\documentclass[a4paper,12pt]{article}

\usepackage[a4paper,left=2.73cm,right=2.7cm,top=3cm,bottom=3.5cm]{geometry}

\usepackage{amsmath,amssymb,mathrsfs,empheq}

\usepackage[colorlinks=true,linktocpage=true,linkcolor=blue,citecolor=blue,urlcolor=blue]{hyperref}

\usepackage[sort&compress,merge,numbers]{natbib}

\addtocontents{toc}{\protect\setcounter{tocdepth}{2}}
\numberwithin{equation}{section}

\RequirePackage[parfill]{parskip}

\usepackage{scalefnt}
\usepackage[titles]{tocloft}
\usepackage{sectsty}
\allsectionsfont{\bf \scalefont{.8} \selectfont}
\subsectionfont{\bf \scalefont{.85} \it \selectfont}
\subsubsectionfont{\bf \scalefont{1} \it \selectfont}

\usepackage{epsfig}
\usepackage{graphicx}
\usepackage{epstopdf}
\usepackage{caption}
\usepackage{subcaption}

\def\BB{{\cal B}}
\def\CC{{\cal C}}

\def\FF{{\cal F}}

\def\JJ{{\cal J}}

\def\LL{{\cal L}}
\def\MM{{\cal M}}
\def\NN{{\cal N}}

\def\QQ{{\cal Q}}
\def\RR{{\cal R}}

\def\TT{{\cal T}}

\def\IR{{\mathbb R}}

\def\IC{{\mathbb C}}

\def\pa{{\partial}}
\def\bS{{\boldsymbol S}}
\def\T{{\boldsymbol T}}

\def\tM{{\tilde M}}

\newcommand{\be}{\begin{equation}}  
\newcommand{\ee}{\end{equation}}
\newcommand{\beq}{\begin{equation}}
\newcommand{\eeq}{\end{equation}}
\newcommand{\bea}{\begin{eqnarray}}
\newcommand{\eea}{\end{eqnarray}}

\DeclareRobustCommand{\DIEP}{\ensuremath{%
\mathchoice{\includegraphics[height=2ex]{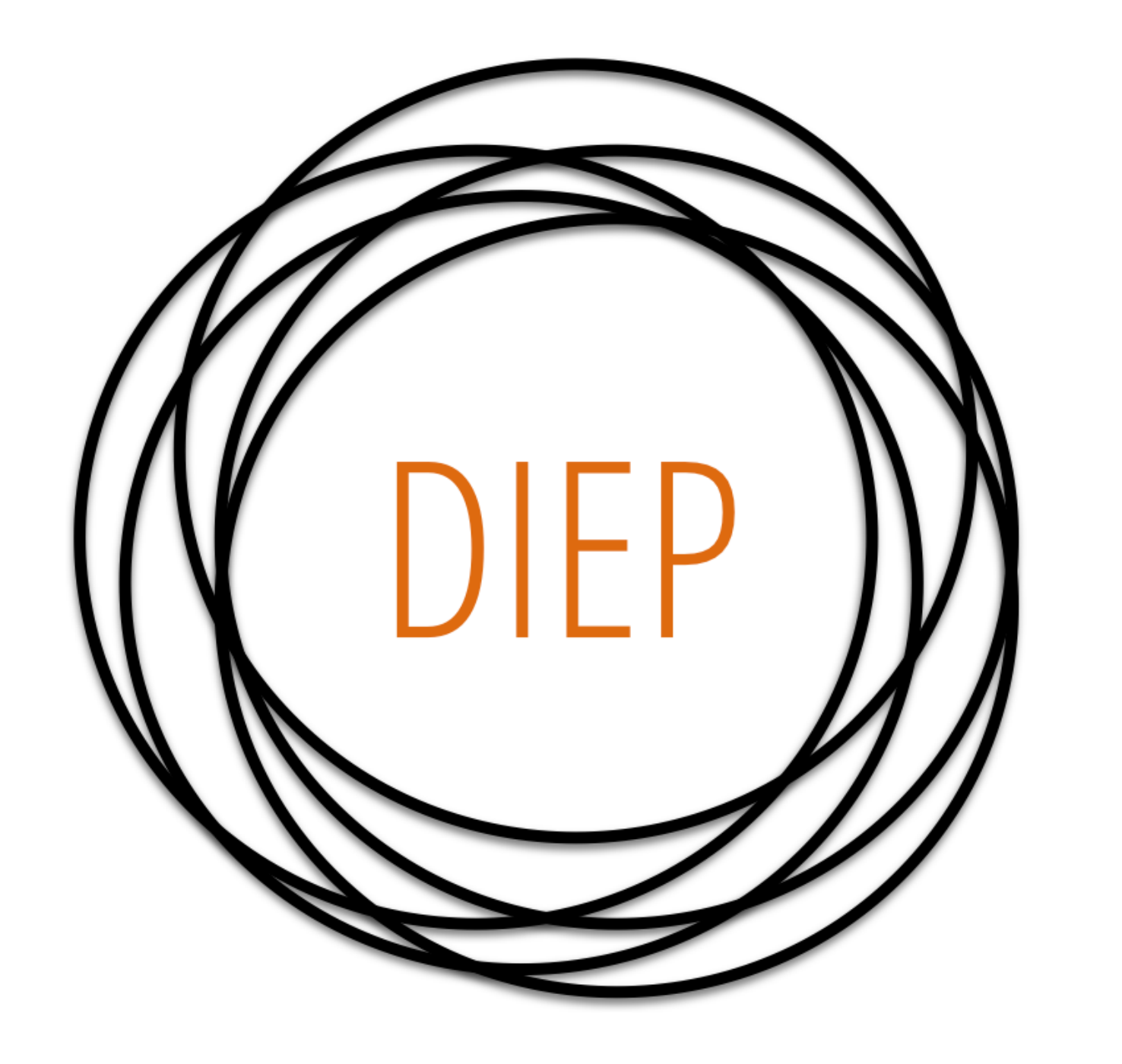}}
  {\includegraphics[height=2ex]{DIEPs.pdf}}
    {\includegraphics[height=1.5ex]{DIEPs.pdf}}
    {\includegraphics[height=1ex]{DIEPs.pdf}}
  }}

\usepackage[nostyling]{jatex}                 

\begin{document}

\begin{titlepage}
\begin{flushright}
DCPT-19/11 \\
NORDITA-2019-039
\vskip -1cm
\end{flushright}

\thispagestyle{empty}

\vspace{25pt}  
	 
\begin{center}
    \centerline{\LARGE
      \bf Thermal transitions of metastable M-branes}
	\vspace{30pt}
		
{\bf Jay Armas$^{\natural,\DIEP}$, Nam Nguyen$^{\dagger}$, Vasilis Niarchos$^{\dagger}$,}
{\bf Niels A. Obers$^{\ddagger,\sharp}$}
		
\vspace{10pt}

{$^\natural$Institute for Theoretical Physics, University of Amsterdam,\\
1090 GL Amsterdam, The Netherlands\\ 
\vspace{5pt}
$\DIEP$ Dutch Institute for Emergent Phenomena, \\
1090 GL Amsterdam, The Netherlands\\
\vspace{5pt}
$^{\dagger}$ Department of Mathematical Sciences and Centre for Particle Theory \\
Durham University, Durham DH1 3LE, United Kingdom\\
\vspace{5pt}
$^\ddagger$
Nordita, KTH Royal Institute of Technology and Stockholm University, Roslagstullsbacken 23, SE-106 91 Stockholm, Sweden \\ 
\vspace{5pt}
$^\sharp$ The Niels Bohr Institute, University of Copenhagen \\
Blegdamsvej 17, DK-2100 Copenhagen Ø, Denmark \\
}

\vspace{10pt}
{\tt 
\href{mailto:j.armas@uva.nl}{j.armas@uva.nl},
\href{mailto:nam.h.nguyen@durham.ac.uk}{nam.h.nguyen@durham.ac.uk},\\
\href{mailto:vasileios.niarchos@durham.ac.uk}{vasileios.niarchos@durham.ac.uk},
\href{mailto:obers@nbi.dk}{obers@nbi.ku.dk}
}

\vspace{20pt}

\abstract{\noindent We use blackfold methods to analyse the properties of putative supergravity solutions in M-theory that describe the backreaction of polarised anti-M2 branes (namely, M5 branes wrapping three-cycles with negative M2-brane charge) in the Cvetic-Gibbons-Lu-Pope background of eleven-dimensional supergravity. At zero temperature we recover the metastable state of Klebanov and Pufu directly in supergravity. At finite temperature we uncover a previously unknown pattern of mergers between fat or thin M5-brane states with the thermalised version of the metastable state. At sufficiently small values of the anti-brane charge a single fat-metastable merger follows the same pattern recently discovered for polarised anti-D3-branes in the Klebanov-Strassler solution in type IIB supergravity. We provide quantitative evidence that this merger is driven by properties of the horizon geometry. For larger values of the anti-brane charge the wrapped M5-brane solutions exhibit different patterns of finite-temperature transitions that have no known counterpart in the anti-D3 system in Klebanov-Strassler.}

\end{center}

\end{titlepage}

\tableofcontents

\begin{center}
\hrulefill
\end{center}
\vspace{-10pt}

\section{Introduction}\label{sec:intro}
\vspace{-0.3cm}

The backreaction of anti-branes in string theory backgrounds with fluxes is a complicated problem with important ramifications. Anti-branes have been used in the past to study supersymmetry breaking in string theory and holography \cite{Maldacena:2001pb, Kachru:2002gs}, to engineer de Sitter solutions \cite{Kachru:2003aw} and study inflationary model building \cite{Kachru:2003sx}, and to construct non-extremal black hole microstates \cite{Bena:2012zi}. The range of applications is large, but the detailed mechanisms of anti-brane backreaction have been the subject of much debated controversies.

\vspace{-0.5cm}
\subsection{Comments on anti-D3 backreaction in Klebanov-Strassler}
\vspace{-0.3cm}

One of the prime examples of anti-brane backreaction in the presence of fluxes involves anti-D3 branes in the Klebanov-Strassler (KS) warped deformed conifold solution in type IIB string theory \cite{Klebanov:2000hb}. This background, which involves non-trivial 3-form NSNS and RR fluxes, is holographically dual to a 4d $\NN=1$ supersymmetric quantum field theory (QFT) with $SU(N)\times SU(N+M)$ gauge group. The  original probe computation by Kachru, Pearson and Verlinde (KPV) in Ref.\ \cite{Kachru:2002gs} revealed that $p$ anti-D3 branes form a metastable state, where the anti-D3s polarise into NS5 branes wrapping an $S^2$ inside an $S^3$ at the tip of the KS throat geometry. The metastable state exists when $p/M$ is sufficiently small, $p/M \lesssim 0.08$.\footnote{$p$ is the number of anti-D3 branes and $M$ the units of 3-form flux through the $S^3$ of the KS geometry.}

The probe computation was performed in \cite{Kachru:2002gs} in two complementary ways: $(i)$ using the worldvolume theory of the anti-D3 branes and $(ii)$ using a worldvolume theory for NS5 branes. In the D3 perspective $(i)$, the non-abelian DBI action is best understood in the super-Yang-Mills limit, which, effectively, restricts the description close to the north pole of the $S^3$. The NS5 brane perspective $(ii)$ does not have this restriction but the formulation of an effective worldvolume theory for NS5 branes is more challenging. KPV employed an abelian DBI action that arises by S-duality from the DBI action of the D5 brane. This step is dubious (see e.g.\ \cite{Bena:2014jaa}), because it is in conflict with the regime of validity of the probe approximation that requires $g_s p\ll 1$ ($g_s$ is the string coupling constant).

The fate of the metastable state beyond the probe approximation involves higher levels of complexity. Considerable effort has been devoted to understand the properties of backreaction in the supergravity regime where one needs to construct backreacted anti-D3 brane solutions with KS asymptotics. Many works, starting with \cite{Bena:2009xk}, revealed solutions of the supergravity equations that involved unphysical singularities in the 3-form fluxes.\footnote{Related earlier work includes \cite{DeWolfe:2008zy,McGuirk:2009xx}. Subsequent developments after \cite{Bena:2009xk} include \cite{Bena:2010ze,Dymarsky:2011pm,Bena:2011hz,Gautason:2013zw,Blaback:2012nf}.} The presence of these singularities was viewed by some authors as evidence that backreaction can change dramatically the conclusions of the probe approximation casting doubt to the very existence of the metastable state originally discovered by KPV (and its subsequent applications to string phenomenology, e.g.\ \cite{Kachru:2003sx}). This conclusion was challenged, however, by the authors of \cite{Michel:2014lva} who argued that the inclusion of backreaction effects in the effective field theory of a {\it single} anti-D3 brane are mild and under control, as one would naively expect.

The non-extremal properties of anti-D3 branes can provide further information about the physics of the system. The thermal properties of anti-D3 black branes in the Klebanov-Strassler background were discussed in a series of papers \cite{Bena:2012ek,Bena:2013hr,Blaback:2014tfa,Hartnett:2015oda}.

In the overwhelming majority of the supergravity constructions the discussion centred around the physics of the backreaction of point-like anti-D3 branes. There are several reasons why the NS5-brane point of view is more appropriate: 
\begin{itemize}
\item[1)] The metastable state in the probe computation of \cite{Kachru:2002gs} is a spherical NS5 state.
\item[2)] A natural candidate for the resolution of the observed supergravity singularities involves the formation of a spherical NS5-brane state \'a la Polchinski-Strassler \cite{Polchinski:2000uf}.
\item[3)] The exact supergravity arguments of \cite{Cohen-Maldonado:2015ssa,Cohen-Maldonado:2016cjh}, which are based on Smarr relations, provide no-go theorems for supergravity solutions describing point-like anti-D3 branes, but leave wide open the possibility of regular spherical NS5 brane solutions. 
\end{itemize}
For all these reasons, a proper understanding of anti-D3 backreaction in the KS background requires information about spherical NS5-brane states. Without this information, previous indications, either from the probe or the supergravity computations with point-like or smeared anti-D3 brane solutions, remain inconclusive.

Finding a supergravity solution that describes spherical NS5 branes with KS asymptotics is a daunting task requiring the solution of complicated systems of partial differential equations in supergravity. Exact analytic solutions to this problem have been notoriously out of reach. For that reason, it is sensible to shift the focus towards more versatile approximation schemes of the supergravity equations. The blackfold formalism \cite{Emparan:2009cs,Emparan:2009at,Armas:2016mes}, which combines elements of matched asymptotic expansions (see e.g. \cite{Harmark:2003yz,Gorbonos:2004uc} for a discussion of matched asymptotic expansions in the context of caged black holes) and higher-form fluid hydrodynamics, has provided very useful information about such problems in the past, starting with \cite{Emparan:2007wm} that explored the properties of neutral higher-dimensional black ring solutions in pure Einstein gravity. All attempts to uncover such solutions in six and higher spacetime dimensions using exact solution generating techniques so far have failed! This is not encouraging for exact approaches, since the case of black rings in flat space is expected to be simpler than the case of wrapped NS5 branes in a geometry with fluxes like the Klebanov-Strassler background.

In recent work \cite{Armas:2018rsy}, we argued that there is a regime of parameters where (extremal and non-extremal) spherical NS5 branes in the KS background should be well approximated by D3-NS5 blackfolds. In this regime, if a full-fledged supergravity solution exists it should obey a certain set of dynamical equations which can be expressed as equations for an effective six-dimensional worldvolume theory. The analysis in  \cite{Armas:2018rsy} revealed the following key points:
\begin{itemize}

\item[$1)$] Extremal spherical NS5 branes should obey at leading order in the blackfold expansion the same equations that were employed by KPV, namely the equations that arise from the S-dual of the DBI action for the D5 brane. Since these equations are now derived directly in the supergravity regime, there is no clash between different regimes of validity and the extremal KPV metastable state can be derived in a long-wavelength approximation directly in gravity. That removes one of the criticisms against the KPV metastable state.

\item[$2)$] In the appropriate regime of parameters, $p/M\lesssim 0.08$, the extremal spherical NS5 branes exhibit two vacua away from the north and south poles: one metastable and one unstable. The blackfold analysis revealed that as soon as the branes become non-extremal an additional unstable vacuum appears. This is a novel, `fat' unstable NS5-brane state. Increasing the entropy of the solutions leads to a merger of the fat unstable state with the thin metastable state. As a function of the entropy, the non-extremal system exhibits a transition with the features of saddle-node bifurcation. Quantitative evidence was provided in \cite{Armas:2018rsy} suggesting that the origin of this transition is closely related to the geometric properties of the corresponding black hole horizons.

\item[$3)$] The emerging picture from the blackfold analysis is suggestively consistent with the exact analysis of \cite{Cohen-Maldonado:2015ssa}. In all cases where \cite{Cohen-Maldonado:2015ssa} lifted a no-go theorem the blackfold approach produced a go with concrete quantitative predictions.
\end{itemize}

\subsection{Outline and summary of results}

The analysis of \cite{Armas:2018rsy} can be extended in several directions. In this paper we explore extremal and finite-temperature metastable configurations of anti-branes in a different background with fluxes. Our goal is two-fold: to test the blackfold approach in more examples of anti-brane backreaction and to supplement known results with new predictions. 

We will focus on another much-studied example of anti-brane backreaction that involves anti-M2 branes in a warped product of $\IR^{2,1}$ and the eight-dimensional Stenzel space in M-theory ---the CGLP background \cite{Cvetic:2000db}. The Stenzel space is a deformed cone over the Sasaki-Einstein manifold $V_{5,2}=SO(5)/SO(3)$, which is the base of the Calabi-Yau 4-fold $\sum_{i=1}^5 z_i^2=0$. Similar to the KPV construction in type IIB string theory, the anti-M2 branes are placed at the tip of the cone, where they backreact in the presence of non-trivial four-form fluxes. This situation was first analysed in the probe approximation by Klebanov and Pufu (KP) in \cite{Klebanov:2010qs}, who found a metastable stable of anti-M2 branes polarised into M5 branes wrapping a three-cycle for $p/\tilde M\lesssim 0.054$ (here $p$ is the number of anti-M2s and $\tilde M$ the units of the background four-form flux). The CGLP background is asymptotically $AdS_4\times V_{5,2}$ and describes holographically a dual three-dimensional gauge theory (see e.g.\ \cite{Martelli:2009ga,Jafferis:2009th,Herzog:2000rz} for aspects of the dual QFT.) The KP vacuum expresses holographically a metastable state in the strongly coupled dual QFT. 

The fate of the KP state beyond the probe approximation was considered in a series of papers. In \cite{Bena:2010gs,Massai:2011vi,Cottrell:2013asa,Giecold:2013pza,Blaback:2013hqa} unphysical four-form flux singularities in supergravity solutions of backreacted anti-M2s were observed casting doubt on the existence of the KP state. In \cite{Cohen-Maldonado:2016cjh} an exact supergravity analysis based on a Smarr formula provided a natural explanation of the observed singularities and indicated that the no-go theorems for point-like anti-M2 branes can be evaded for spherical M5 branes in agreement with the probe computation. 

The no-go theorems of \cite{Cohen-Maldonado:2016cjh} make the following predictions about exact supergravity solutions of anti-M2 branes in the CGLP background:
\begin{itemize}
\item[1)] There are no regular extremal solutions of point-like anti-M2 branes. Here and in what follows, the terminology `point-like anti-M2' refers to solutions with vanishing M5 brane dipole charge and spherical horizon topology (more precisely, $\IR^2 \times S^7$ horizon topology). Regular solutions of M5 branes wrapping a 3-cycle, with horizon topology $\IR^2 \times S^3 \times S^4$, can evade the no-go theorem.
\item[2)] At finite temperature, point-like anti-M2 black brane solutions can in principle exist, but require a specific boundary condition for the gauge potentials at the horizon. As we noted in the previous point, they cannot have a regular extremal limit. Regular black M5 brane solutions wrapping a 3-cycle are in principle allowed. 
\end{itemize}

As a result, we expect that as we raise the temperature, the horizon size of a putative metastable M5 black brane state will increase eventually eating up the $S^3$ part of the geometry and converting the horizon topology to an $S^7$. At that point (or close to that point), either the solution is one of the spherical horizon topology solutions (if such solutions exist) or the metastable black M5 state disappears. Whatever the outcome, this picture suggests a finite temperature transition. 

One can imagine different mechanisms by which a metastable M5 state disappears in a scenario where it cannot transit to a solution of spherical horizon topology. One possibility is a mechanism driven by horizon geometry. Near the critical point, the deformations of the horizon geometry play an important role and lead to the loss of the metastable state before the solution changes topology. Another possibility is that the metastable state disappears before the size of the Schwarzschild radius grows significantly because of modifications of the finite-temperature potential that resemble how the metastable state is lost at zero temperature at sufficiently high values of the anti-charge.

In this paper we uncover, using the blackfold formalism, a black hole phase diagram that is not only consistent with these expectations and the no-go theorems of \cite{Cohen-Maldonado:2016cjh}, but also reveals new unexpected patters of finite-temperature transitions. Interestingly, we find (unlike the case of anti-D3 branes in Klebanov-Strassler) that both of the possibilities mentioned in the previous paragraph can appear in different regimes of parameters.

At zero temperature, we show that the blackfold equations recover faithfully the abelian DBI equations used by KP in \cite{Klebanov:2010qs} and the same extremal metastable vacuum that they found. Alternatively, these equations could be obtained with the use of the abelian PST effective action for M5 branes \cite{Pasti:1996vs,Pasti:1997gx,Bandos:1997ui}. The fact that the blackfold equations for extremal M2-M5 brane bound states are closely related to the equations of motion of the PST effective action has also been noted previously in \cite{Niarchos:2015moa,Niarchos:2014maa}. In section \ref{extremal} we re-affirm this statement. The main lesson of this part is that the KP metastable state is consistent with the constraint equations of supergravity for a putative wrapped M5 brane configuration. As we note in section \ref{constraintequations}, this is a strong indication that one can setup a matched asymptotic expansion scheme to obtain perturbatively a regular solution of fully backreacted metastable wrapped M5 branes in supergravity.

At finite, sufficiently small, temperature, we uncover (in direct analogy to the case of anti-D3 branes in Klebanov-Strassler \cite{Armas:2018rsy}) three main branches of wrapped M5 black brane solutions: a fat unstable state, a metastable state and a thin unstable state. The terms `fat' and `thin' refer to the relative size of the $S^3$ that the M5 brane wraps and the size of the Schwarzschild radius. The behaviour of these branches at higher temperatures depends on the value of $p/\tilde M$. Surprisingly, in section \ref{meta} we discover three separate regimes of $p/\tilde M$ (inside the window of the metastable state, $p/\tilde M \lesssim 0.054$) that exhibit different patterns of thermal transitions.

There is a low-$p/\tilde M$ regime where the anti-M2 physics in CGLP is very similar to the anti-D3 physics in Klebanov-Strassler. In this regime there is a single finite-temperature transition that involves the merger of a fat unstable black M5 with the metastable black M5. Beyond this merger the metastable state is lost. We present non-trivial quantitative evidence that supports the scenario where this merger is driven by properties of the horizon geometry. This observation gives us confidence that the leading order blackfold equations capture quite accurately the long-wavelength properties of full-fledged black hole solutions. 

In addition, for the anti-M2 system we find two regimes of $p/\tilde M$ that have no known counterpart in the system of anti-D3 branes in Klebanov-Strassler. In the large-$p/\tilde M$ regime there is a single merger between the metastable state and the {\it thin} unstable M5 brane state. In this case, there are no indications that the loss of the metastable state is driven by properties of the horizon geometry. In an intermediate regime of $p/\tilde M$ the phase diagram exhibits three (instead of one) transitions: two of them involve mergers of the metastable state with the thin unstable state and one involves a merger of the metastable state with the fat unstable state. These patterns are new, unexpected predictions of the blackfold formalism for supergravity and the dual QFT. 

The plan of the paper is as follows. Useful properties of the CGLP background are reviewed in section \ref{CGLP}. The key components of the formalism that we use and the regime of our approximations are discussed in section \ref{blackfolds}. The recovery of the extremal KP state from the blackfold equations is presented in section \ref{sec:extremal}. The non-extremal blackfold equations and several related types of effective potentials that facilitate different aspects of our analysis are discussed in section \ref{thermal}. The main results on the non-extremal properties of the system are obtained in section \ref{meta}. Important questions and open problems are summarised in the concluding section \ref{conclusions}. A note on Smarr relations is relegated to appendix \ref{smarr}.

\section{M-theory setup}
\label{CGLP}


We consider the backreaction of polarised anti-M2 branes in the eleven-dimensional supergravity solution of CGLP \cite{Cvetic:2000db}, which is a warped product of $\IR^{2,1}$ and the eight-dimensional Stenzel space. The eleven-dimensional background metric has the form
\beq
\label{setupaa}
ds_{11}^2 = g_{\mu\nu} dx^\mu dx^\nu = H^{-\frac{2}{3}} \left( - (dx^0)^2 +  (dx^1)^2 +  (dx^2)^2 \right)
+ H^{\frac{1}{3}} ds_8^2
~,
\eeq
where $ds_8^2$ is the metric element of the Stenzel space. Details on the full structure of this metric and the function $H$ can be found in \cite{Cvetic:2000db,Klebanov:2010qs}. The background also involves a non-trivial profile for the four-form field strength $G_4$ and its Hodge dual $G_7 = \star_{11} G_4$. It is convenient to collect the following constants that appear in this background (we follow closely the notation in \cite{Klebanov:2010qs}) 
\begin{itemize}

\item A constant $m$ appears in the expressions of $G_4$ and $G_7$ and is related to the $\tM$ units of $G_4$ flux through an $S^4$ of the background via the relation
\beq
\label{setupab}
\tM = \frac{18\pi^2 m}{(2\pi \ell_P)^3}
~,
\eeq
where $\ell_P$ is the eleven-dimensional Planck length.

\item The complex structure deformation of the four-complex dimensional conifold that gives rise to the Stenzel space is expressed in terms of the constant $\epsilon$. In complex coordinates $z_i$ in $\IC^5$ the Calabi-Yau space $\sum_{i=1}^5 z_i^2 = 0$ is deformed to  $\sum_{i=1}^5 z_i^2 = \epsilon^2$.

\item At the tip of the cone the value of the function $H$ is 
\beq
\label{setupac}
 \hat H_0 \frac{m^2}{\epsilon^{\frac{9}{2}}} \simeq 1.0898\, \frac{m^2}{\epsilon^{\frac{9}{2}}}
~.
\eeq

\item It is useful to consider the related constants
\beq
\label{setupad}
a_0^2 = \left( \frac{m^2}{\epsilon^{\frac{9}{2}}} \hat H_0 \right)^{-\frac{2}{3}}~, ~~
b_0^2 = \frac{3}{2} \hat H_0^{\frac{1}{3}}
~.
\eeq

\end{itemize}

Anti-M2 branes placed in the CGLP background are attracted towards and eventually stabilise at the tip of the eight-dimensional cone. Hence, for our purposes it is enough to focus on the tip of the conifold ($\tau = 0$ in the appropriate radial coordinate $\tau$ \cite{Klebanov:2010qs}). After a trivial rescaling of the Minkowski coordinates $x^0, x^1, x^2$ by the constant factor $a_0 /({b_0 m^{1/3}})$ one obtains the metric\footnote{At the tip, we localise at the origin of the transverse four-dimensional flat space. Hence, in \eqref{setupae} only an $S^4$ out of the overall seven-dimensional transverse space appears.}
\beq
\label{setupae}
ds^2 = m^{\frac{2}{3}} b_0^2 \left( - (dx^0)^2 +  (dx^1)^2 +  (dx^2)^2 + d\psi^2 + \sin^2\psi \, d\Omega_3^2 \right)
~.
\eeq
   We will use spherical coordinates $\vartheta, \omega, \varphi$ to express the metric element of the unit round $S^3$ as $d\Omega_3^2 = d\vartheta^2 + \sin^2\vartheta \left( d\omega^2 + \sin^2\omega \,d\varphi^2 \right)$. The four-form flux $G_4=d A_3$ is given in terms of the gauge field
\beq
\label{setupaf}
A_3 = \frac{27}{4} m f(\psi) \sin^2 \vartheta \sin\omega \, d\vartheta \wedge d\omega \wedge d\varphi
~,
\eeq
where
\beq
\label{setupag}
f(\psi) = \frac{1}{3} \cos^3 \psi - \cos\psi + \frac{2}{3}
~,
\eeq
while the seven-form flux $G_7$ takes the form
\beq
\label{setupai}
G_7 = - \frac{27}{4} m^2 b_0^3 \sin^3\psi  \sin^2\vartheta \sin\omega \, dx^0 \wedge dx^1 \wedge dx^2 \wedge d\psi \wedge d\vartheta \wedge d\omega \wedge d\varphi
~.
\eeq

\section{Forced M-brane blackfolds and higher-form hydrodynamics}
\label{blackfolds}

We are interested in supergravity solutions that describe the polarization of M2 branes into M5 branes wrapping an $S^3$ inside the $S^4$ of the background geometry \eqref{setupae}. As we noted in the introduction there is almost no information about such solutions in the current literature, because the exact solutions of this type involve the analysis of intractable partial differential equations. 

A more efficient approach involves a long-wavelength expansion scheme of the supergravity equations where one tries to match a solution that interpolates between the asymptotic background \eqref{setupae}-\eqref{setupai} at large distances and the near-horizon region of a wrapped M2-M5 bound state solution. In the appropriate regime of parameters, which will be discussed in detail later in this section, the local behaviour of the latter can be further approximated by the M2-M5 bound state solution in flat space. For quick reference, the M2-M5 bound state in flat space has metric \cite{Izquierdo:1995ms} (see also \cite{Harmark:2000ff,Harmark:1999rb})
\begin{align}
\begin{split}
\label{forceaa}
ds^2 &= (HD)^{-\frac{1}{3}} \left[ - f (dx^0)^2 + (dx^1)^2 +(dx^2)^2 + D( (dx^3)^2+(dx^4)^2+(dx^5)^2)\right]
\\
&~~+ H \left( f^{-1} dr^2 + r^2 d\Omega_4^2 \right)
~,
\end{split}
\end{align}
where
\beq
\label{forceab}
f = 1 - \frac{r_0^3}{r^3}~, ~~ H = 1+\frac{r_0^3 \sinh^2\alpha}{r^3}~, ~~
D = (\sin^2\theta \, H^{-1} + \cos^2\theta )^{-1}
~.
\eeq
There are also non-trivial profiles for the 3-form potential $A_3$ and the dual potential $A_6$
\beq
\label{forceac}
A_3 = - \sin\theta \coth\alpha \, (H^{-1}-1) \, dx^0 \wedge dx^1 \wedge dx^2 + \tan\theta \, D H^{-1} dx^3 \wedge dx^4 \wedge dx^5
~,
\eeq
\beq
\label{forcead}
A_ 6= \cos\theta \, \coth\alpha \, D (H^{-1}-1) dx^0 \wedge dx^2 \wedge \cdots \wedge dx^5
~.
\eeq
The M5 worldvolume directions are along $(012345)$ and there is a smeared density of M2-brane charge along the directions $(012)$. The parameters $r_0,\alpha,\theta$ are parameters of the non-extremal solution. They parametrise the thermodynamic quantities of the solution (the energy density $\varepsilon$, the temperature $\TT$, the entropy density $s$, the M2 and M5 chemical potentials $\Phi_2,\Phi_5$ and the corresponding charge densities $\QQ_2, \QQ_5$). In the Einstein frame we have the following relations
\beq
\label{forceaea}
\varepsilon = \frac{\Omega_4}{16\pi G} r_0^3 (4+3\sinh^2\alpha)~, ~~ \TT=\frac{3}{4\pi r_0 \cosh\alpha}
~,
\eeq
\beq
\label{forceaeb}
s = \frac{\Omega_4}{4G} r_0^4 \cosh\alpha ~, ~~ \Phi_2 = - \sin\theta \tanh\alpha~, ~~ \Phi_5 = \cos\theta \tanh\alpha
~,
\eeq
\beq
\label{forceaec}
\QQ_2 = - 
\frac{3\Omega_4}{16\pi G} r_0^3 \sin\theta  \sinh\alpha \cosh\alpha~, ~~
\QQ_5 = \frac{3\Omega_4}{16\pi G} r_0^3 \cos\theta \sinh\alpha \cosh\alpha
~,
\eeq
where $\Omega_4=\frac{8\pi^2}{3}$ is the volume of the unit 4-sphere $S^4$ and $G=(2\pi)^8 \ell_P^9$ is Newton's gravitational constant. For later reference we also define the quantity
\beq
\label{forceaeca}
\widetilde \QQ_ 2 = \frac{3\Omega_4}{16\pi G}  r_0^3 \sin\theta \cos\theta \sinh^2\alpha 
~,
\eeq
which is obtained from the solution \eqref{forceaa}-\eqref{forcead} from an integral of the four-form field strength $G_4=dA_3$ over an $S^7$  that surrounds the worldvolume directions that are orthogonal to the dissolved M2-brane density: $\widetilde \QQ_2 = -\frac{1}{16\pi G} \int_{S^7} \star G_4$. The expression of the corresponding free energy $\FF$ (which will be useful later on) is
\beq
\label{forceaed}
\FF = \varepsilon - \TT s =  \frac{\Omega_4}{16\pi G} r_0^3 ( 1+ 3\sinh^2\alpha)
~.
\eeq
The extremal limit is achieved by taking $r_0\to 0$, $\alpha\to \infty$ with the combination $r_H^3 \equiv r_0^3 \sinh^2\alpha$ kept fixed.

When the values of the parameters $r_0$ and $\theta$ are non-zero the $SO(5,1)$ Lorentz symmetries of the M5-brane worldvolume are spontaneously broken by the temperature and the presence of the M2-brane density. The Goldstone bosons of these symmetries can be phrased in terms of a unit time-like velocity vector $u^a$ and a projector $\hat h_{ab}$ which is aligned along the worldvolume directions of the smeared M2-brane density inside the M5.\footnote{In full generality, Goldstone scalars are expected to be required to be introduced to describe the dynamics of the system as for smeared string density \cite{Armas:2018atq, Armas:2018zbe}. However, for the cases studied in this paper, it is sufficient to work with $\hat h_{ab}$ .} From here on, latin letters from the beginning of the alphabet $a,b,\ldots=0,1,\ldots, 5$ will be used to denote the M5-brane worldvolume directions. $\hat h_{ab}$ can be further expressed in terms of three unit spacelike orthonormal worldvolume vectors $v^a,w^a,z^a$ (which are also normal to $u^a$) 
\beq
\label{forceaf}
\hat h_{ab} = \eta_{ab} - v_a v_b - w_a w_b - z_a z_b
~~,
\eeq
where $\eta_{ab}$ is the 6d Minkowski metric. In terms of these variables the worlvolume energy-momentum tensor takes the form
\beq
\label{forceag}
T_{ab} = \TT s \left( u_a u_b - \frac{1}{3} \eta_{ab} \right) - \Phi_2 \QQ_2 \hat h_{ab} - \Phi_5 \QQ_5 \eta_{ab}
~.
\eeq

With these ingredients one would like to set up a scheme of matched asymptotic expansions where the supergravity equations are solved perturbatively. In this scheme a wrapped M5-brane solution is captured at leading order in the near horizon region by a long-wavelength deformation of the M2-M5 bound state \eqref{forceaa}-\eqref{forcead} where the parameters $r_0,\theta$, $\hat h_{ab}$ are promoted to slowly varying functions of the M5 worldvolume directions $\sigma^a$ $(a=0,1,\ldots,5)$. The extrinsic bending of the worldvolume in the ambient eleven-dimensional spacetime is captured by promoting the trivial Minkowski metric $\eta_{ab}$ in \eqref{forceag} into a non-trivial slowly varying induced metric $\gamma_{ab}$.

Let us denote by $r_b$ the characteristic scale of the near-horizon solution, by $L$ the characteristic scale of the asymptotic background and by $\RR$ the characteristic scale of the above worldvolume quantities. Typically, the scales $L$ and $\RR$ are not completely unrelated because a non-trivial background forces a solution to develop comparable gradients. We would like to work in a regime where $r_b \ll \min(\RR,L)$. The small ratio $r_b/\RR \ll 1$ guarantees that we can set up a matched asymptotic expansion where a {\it near-zone} solution at transverse distances $r\ll \RR$ can be matched to a {\it far-zone} solution at $r\gg r_b$ on a large {\it overlap} region $r_b \ll r \ll \RR$. The simultaneously small ratio $r_b/L \ll 1$ guarantees that the leading order near-zone solution can be approximated by the (black) M2-M5 brane solution in {\it flat space}. Concrete examples of this expansion in AdS spacetimes can be found in \cite{Caldarelli:2008pz,Armas:2010hz} (a related general discussion appears in section 5 of \cite{Armas:2016mes}). The specifics of the conditions imposed by these small ratios in our setup will be presented in subsection \ref{validity} below.

In general cases, obtaining solutions of long-wavelength deformations in the near horizon region is highly non-trivial even at the leading order of the expansion. The existence of regular solutions was worked out systematically in pure Einstein gravity in \cite{Camps:2012hw} and in Einstein-Maxwell-dilaton theories in \cite{Gath:2013qya,DiDato:2015dia}. It remains an open problem in general supergravity theories for multi-charge solutions, though the set of necessary constraint equations that need to be solved for the existence of such solutions has been identified in full generality \cite{Armas:2016mes}, as will be explained below. Despite the lack of a general complete construction of regular solutions in such matched asymptotic expansions in supergravity, the analysis of the above cases and the many successful applications of the blackfold formalism has led to the appreciation that one can distill very useful, immediate information about a putative gravitational solution by focusing the analysis on the constraint equations of (super)gravity and studying them independently of the rest of the (super)gravity equations.

\subsection{On the dominant role of the constraint equations}
\label{constraintequations}

The (super)gravity constraint equations for (black) brane ans\"atze of the above-mentioned type are conservation equations of effective currents (the energy-momentum tensor and other higher-form currents) that can be identified either in the near-zone or the far-zone expressions of the perturbative solution.\footnote{For supersymmetric M2-M5 solutions in flat space it has been shown \cite{Niarchos:2014maa} that one can also find an analog of the usual gravitational constraint equations in the Killing spinor equations. These equations coincide with the $\kappa$-symmetry conditions of the abelian M5-brane theory.}  In the leading order far-zone analysis they become the well-known (forced) conservation equations of the Newtonian-approximation. The importance of the constraint equations relies on three central points:
\begin{itemize}
\item[$(i)$] They are necessary conditions for the existence of a perturbative solution. This is most evident in the asymptotic Newtonian analysis of the matched asymptotic expansion.
\item[$(ii)$] In all cases that have been worked out in detail (most notably  \cite{Camps:2012hw} in Einstein gravity) the constraint equations are also sufficient conditions for the existence of a regular solution. Specifically, the solution of the leading order constraint equations guarantees the consistent solution of {\it all} the gravitational equations at first order. The expectation that this is true to all orders of the matched asymptotic expansion in general theories of gravity and for general black brane solutions was dubbed `the {\it blackfold conjecture}' in \cite{Niarchos:2015moa}. This conjecture has a closely related cousin in the fluid-gravity correspondence in AdS/CFT \cite{Bhattacharyya:2008jc}. 
\item[$(iii)$] The solution of the leading order constraint equations provides complete thermodynamic information about the first-order backreacted supergravity solution. This is also analogous to the fluid-gravity correspondence (e.g. \cite{Bhattacharyya:2007vs}).\footnote{See also \cite{Emparan:2007wm} for a concrete exhibition of this statement.}
\end{itemize}
The constraint equations are equations of an effective worldvolume theory. Since they capture long-wavelength properties of broken symmetries they are naturally formulated as generalized higher-form hydrodynamics on dynamical hypersurfaces. We will be referring to these equations as {\it `blackfold equations'} and they will be laid out explicitly for the system of interest in the following subsection. 

We would like to stress the following point. The fact that the leading order blackfold equations can be obtained  from an asymptotic Newtonian analysis may give the impression that these equations are describing solutions in a probe approximation where backreaction effects are not taken into account. This viewpoint is misleading. Because of points $(ii)$ and $(iii)$ the blackfold equations capture data of bona fide backreacted supergravity solutions. A clean example of this statement can be found in \cite{Emparan:2007wm} where neutral ultra-spinning thin black rings in higher-dimensional Einstein gravity were analysed using these very same blackfold and matched asymptotic expansion techniques. 

It has been noted \cite{Niarchos:2015moa,Grignani:2016bpq,Armas:2016mes} that by applying the leading order blackfold equations to extremal multi-charge D-brane configurations in ten-dimensional type IIA/B supergravity one can re-discover from supergravity the familiar abelian DBI description of D-branes and all their bulk-boundary couplings in open string theory.\footnote{The DBI action is derived in weakly coupled open string theory when D-branes are treated in the probe approximation. In \cite{Niarchos:2015moa} it was proposed that the simultaneous emergence of the abelian DBI in the opposite, supergravity regime, is a manifestation of the supergravity/DBI correspondence and that its origin should be traced back to open-closed string dualities that generalise the AdS/CFT correspondence beyond the near-horizon decoupling limit. The relation of the leading order blackfold equations with DBI dynamics has been shown to exist at extremality irrespective of supersymmetry \cite{Niarchos:2015moa}.} By studying higher-order corrections of the blackfold equations in the double $r_b/L$, $r_b/\RR$ expansion in the presence of higher-form charges one can obtain interesting modifications of the abelian DBI actions that have not been explored systematically to-date (for previous studies of higher-derivative corrections see \cite{Armas:2011uf,Armas:2012ac,Armas:2012jg,Armas:2013hsa,Armas:2013goa, Armas:2018ibg}).

The blackfold equations extend easily beyond the case of D-branes. At extremality, they also provide DBI-like descriptions for NS5 branes\footnote{The 6d blackfold effective action for NS5 branes is the natural non-linear completion of the NS5 worldvolume action proposed several years ago in \cite{Callan:1991ky}. As we noted in the introduction, this worldvolume description of NS5 branes played a key role in \cite{Armas:2018rsy} where the KPV results were recovered.} as well as M-branes in M-theory whose microscopic derivation is less straightforward. For M5 branes one expects to recover the PST effective action \cite{Bandos:1997ui} in this manner. We will verify this expectation for a special configuration of wrapped M5s in section \ref{sec:extremal} and section \ref{potentials}.

Motivated by points $(i)$, $(ii)$, $(iii)$ above, the main purpose of this paper is to analyse the constraint equations for polarised black anti-M2 branes in the CGLP background. Assuming the validity of $(ii)$ we will be able to obtain in this manner new information about the existence of fully-backreacted metastable states in M-theory in the supergravity regime and data about their finite-temperature thermodynamic properties that have been to-date inaccessible to all other methods. 

The validity of point $(ii)$ is crucial in this exercise. We expect it to be true for the following reasons. Firstly, as we mentioned above there are classes of examples where at least the leading order of the required matched asymptotic expansions work in accordance with $(ii)$. There is a proof of this statement for neutral black holes in Einstein gravity \cite{Camps:2012hw} and in Einstein-Maxwell-dilation theories and related ones \cite{Armas:2013aka}.

For anti-branes in background fluxes one may be especially worried that $(ii)$ may fail. However, by following the exact analysis of \cite{Cohen-Maldonado:2015ssa,Cohen-Maldonado:2016cjh} one can derive specific no-go theorems, where it is obvious when unphysical singularities may occur. The polarised anti-D3s in Klebanov-Strassler and the polarised anti-M2s in CGLP evade these no-go theorems. Although, this does not prove $(ii)$ it provides a very suggestive context where it can be considered. In fact, as we emphasised in the introduction, we can do a little better. 

The exact analysis of \cite{Cohen-Maldonado:2015ssa,Cohen-Maldonado:2016cjh} emphasises the role of horizon topology and makes a natural prediction. On general grounds, one expects that as one increases the temperature of a black hole solution with non-spherical horizon topology, there will be a point, at sufficiently high temperature, where the horizon topology of the solution will change and will become spherical. As we noted in the introduction, this is a natural scenario consistent with the no-go theorems, where the topology change does not occur, because a dramatic change of the horizon geometry is accompanied with the loss of the metastable branch. In section \ref{meta} we show that, according to the blackfold analysis, there is a regime of $p/\tilde M$ where this scenario is indeed verified. In addition, the blackfold equations produce concrete quantitative data that support the claim that this transition is driven by properties of the horizon. In our opinion, this is non-trivial evidence that fits well with the proposed validity of the assumption in point $(ii)$.

It would be very interesting to show $(ii)$ explicitly by constructing the full first-order backreacted supergravity solution. This is a rather complicated task to which we hope to return in future work. Nevertheless, we stress again that by solving all the supergravity equations at the leading order of the matched asymptotic expansion we would demonstrate $(ii)$ (and therefore the perturbative existence of the backreacted polarised anti-M2-branes), but we would not obtain any new information about thermodynamics at this order beyond the data we extract in the present paper.

\subsection{Blackfold equations}
\label{blackequations}

We can summarise the M2-M5 thermodynamics \eqref{forceaea}-\eqref{forceag} with the use of the following currents, namely, the energy-momentum tensor
\begin{equation}
\begin{split}
\label{blackeqsaa}
T_{ab} &= \TT s \left( u_a u_b - \frac{1}{3} \gamma_{ab} \right) - \Phi_2 \QQ_2 \hat h_{ab} - \Phi_5 \QQ_5 \gamma_{ab}
\\
&= \CC \Bigg[ r_0^3 \left( u^a u^b - \frac{1}{3} \gamma^{ab}   \right) 
- r_0^3 \sinh^2 \alpha \gamma^{a b} + r_0^3  \sin^2 \theta \sinh^2 \alpha \left(  v^a v^b + w^a w^b + z^a z^b \right)  \Bigg]
\end{split}
\end{equation}
and the charge currents
\begin{equation}
\begin{split}
\label{blackeqsab}
J_3 &= \QQ_2 * (v \wedge w \wedge z ) - \widetilde \QQ_2 \, v \wedge w \wedge z 
\\
&=  \CC r_0^3 \sin\theta \sinh\alpha \bigg[- \cosh\alpha * (v \wedge w \wedge z ) 
- \cos\theta \sinh\alpha \, v \wedge w \wedge z \bigg]
~,
\end{split}
\end{equation}
\bea
\label{blackeqsac}
\JJ_6 = - \QQ_5 * 1 
= - \CC r_0^3 \cos\theta \sinh\alpha \cosh\alpha * 1 
~.
\eea
In these expressions we introduced for convenience the constant $\CC=\frac{3\Omega_4}{16\pi G} = \frac{\pi}{2G}$. $\gamma_{ab}=g_{\mu\nu} \frac{\partial X^\mu}{\partial \sigma^a} \frac{\partial X^\nu}{\partial \sigma^b}$ is the six-dimensional induced metric on the effective M5-brane worldvolume and $g_{\mu\nu}$ is the eleven-dimensional metric on the asymptotic background.  The scalars $X^\mu$ capture covariantly the embedding of the 6d worldvolume in the ambient background geometry. The Hodge dual $*$ refers to the induced metric $\gamma_{ab}$. We draw the attention of the reader to the two contributions that appear on the RHS of eq.\ \eqref{blackeqsab}. The first contribution is an electric current for the dissolved M2-brane charge and the second a magnetic current that follows from fluxes that are inherent in the M2-M5 bound state.

The general leading order blackfold equations in M-theory in the presence of background fluxes were derived in \cite{Armas:2016mes}. They can be written in the form
\begin{empheq}[box=\widefbox]{align}
\label{blackeqsad}
&\nabla_a T^{a\mu} = \frac{1}{3!} G_4^{\mu a_1 a_2 a_3} J_{3a_1 a_2 a_3} + \frac{1}{6!} G_7^{\mu a_1 \cdots a_6} \JJ_{6a_1\cdots a_6}
~,\\
\label{blackeqsae}
&d \star J_3 + \star \JJ_6 \wedge G_4 = 0
~,\\
\label{blackeqsaf}
&d \star \JJ_ 6 =0
~.
\end{empheq}
In these expressions $\star$ is the Hodge dual with respect to the eleven-dimensional asymptotic background metric $g_{\mu\nu}$.

The combination of eqs.\ \eqref{blackeqsad}-\eqref{blackeqsaf} and \eqref{blackeqsaa}-\eqref{blackeqsac} can be viewed as an effective 6d hydrodynamic system for a perfect fluid of higher-form symmetries on a dynamical worldvolume. They are dynamical equations for the unknown worldvolume functions $r_0,\theta,\alpha, u^a, v^a, w^a, z^a$ and the transverse part of the embedding scalars $X^\mu$.

The current conservation equations \eqref{blackeqsae}, \eqref{blackeqsaf} have a straightforward meaning. Eq.~\eqref{blackeqsaf} implies that the total number of M5 branes is fixed, namely that the quantity $\QQ_5$ is a constant of motion. $\QQ_5$ is a bona fide charge when the M5 brane does not wrap compact surfaces, otherwise it is a dipole charge. The modified conservation equation \eqref{blackeqsae} implies the conservation of an effective current $d \star \tilde J_3=0$ such that the M2-brane charge
\beq
\label{blackeqsag}
\mathbb Q_2 \equiv \int_{\MM_3^\perp} * \tilde J_3 ~~,~~\tilde J_3=J_3 + \star ( \star \JJ_6 \wedge A_3 )~~,
\eeq
is also a constant of motion. In this expression the integral is performed over the worldvolume directions $\MM_3^\perp$ perpendicular to the worldvolume directions $\MM_3$ of the dissolved M2-brane charge. $\mathbb Q_2$ is the Page charge of the M2 branes \cite{Marolf:2000cb}.

\subsection{Regimes of validity}
\label{validity}

The blackfold expansion is a long-wavelength expansion whose validity requires that higher-order corrections are much smaller than the ideal order quantities. Determining the validity of blackfold approximations requires a quantitative understanding of the length scales associated to variations of leading order structures, such as that associated with the curvature of the worldvolume. Many of the geometric structures that determine the scales $L,\mathcal{R}$ discussed above have been analysed in \cite{Armas:2013hsa, Armas:2015kra}, such as the worldvolume Ricci scalar, the spacetime Ricci scalar or the square of the mean extrinsic curvature of the worldvolume. Thus, we must require that $r_b\ll \text{min}(L,\mathcal{R})$ where $r_b$ can be one of the three different length scales characterising the near-horizon brane solution, namely the radii associated with the energy density and charge densities (see \eqref{forceaea}-\eqref{forceaec}) 
\beq
r_b=\left(\frac{\varepsilon}{\mathcal{C}},\frac{\mathcal{Q}_5}{\mathcal{C}}, \frac{\mathbb{Q}_2}{\mathcal{C}}\right)^{1/3}~~,
\eeq
where we recall that $\mathcal{C}=\frac{\pi}{2G}$.

In the following sections we will be considering configurations of M5 branes that are extended along the directions $x^0, x^1, x^2$ and wrap the $S^3$ at fixed angle $\psi$ in the CGLP background \eqref{setupae}. It is then natural to focus our attention on the validity of these specific configurations. The smallest curvature scale is that associated with the curvature of the worldvolume ($\mathcal{R}\sim b_0 m^{1/3}\sin\psi$) and the largest intrinsic scale is the energy density radius ($r_b\sim r_0\sinh\alpha$). Thus we must require
\beq \label{eq:reqm}
r_0\sinh\alpha\ll b_0 m^{1/3}\sin\psi~~,
\eeq
which constrains the product $r_0\sinh\alpha$. It is clear from this expression that the validity of the approximation breaks down both near the north and south poles. The same conclusion is reached if charge density radii are considered. In particular, expressing the M5 brane charge in terms of the number $N_5$ of M5 branes such that
\beq
\mathcal{Q}_5=\frac{N_5}{(2\pi)^5\ell_P^6}~~,
\eeq
and using $16\pi G=(2\pi)^8\ell_P^9$, together with \eqref{setupab}, one must have that
\beq
\left(\frac{N_5}{\tM }\right)^{1/3}\ll b_0\sin\psi~~,
\eeq
which again breaks down near the north and south poles. On the other hand, for values of $\sin\psi\sim1$ one obtains 
\beq
\frac{N_5}{\tM }\ll1~~.
\eeq
When approaching the poles, i.e. for values of $\sin\psi\sim0$, one requires an even smaller ratio of $N_5/\tM$. From the requirement that charge density associated to $\mathbb{Q}_2$ is small enough, one finds 
\beq \label{eq:reqq2}
\left(\frac{p}{\tM}\right)^{1/3}\left(\frac{N_5}{\tM }\right)^{1/3}\ll b_0\sin\psi~~,
\eeq
where we have defined 
\beq
\label{extrai}
\frac{p}{\tilde M}  \equiv \frac{\mathbb Q_2}{18\pi^2 m \QQ_5}
~.
\eeq
From the requirement \eqref{eq:reqq2} we deduce that the ratio $p/\tM$ cannot be too large. We thus conclude that, except very near $\psi=0,\pi$, there is always a choice of parameters that allows for these configurations to be within the regime of validity of the approximation.

\section{KP metastability from extremal blackfolds}
\label{sec:extremal}

As a useful warmup, we consider first the case of a special time-dependent ansatz of wrapped M5 branes. We assume that the conditions of the blackfold expansion are met and that the time-dependence is characterised by appropriately small derivatives. Since we wrap the M5s around the $S^3$ at $\tau=0$ in \eqref{setupae} there is a single transverse scalar that we turn on ---the angle $\psi$. We work in static gauge and set 
\beq
\label{extraa}
x^0 =\sigma^0 \equiv t ~, ~~ x^i =\sigma^i ~(i=1,2)~, ~~ 
\vartheta = \sigma^3~, ~~ \omega = \sigma^4~, ~~ \varphi = \sigma^5
 ~.
 \eeq
 We are taking the extremal limit $r_0\to 0$, $\alpha\to \infty$ keeping the combination $r_H^3 = r_0^3 \sinh^2\alpha$ fixed. In our ansatz there are three unknown functions
 \beq
 \label{extrab}
 \psi(t)~,~~ r_H(t)~, ~~ \theta(t)
 ~.
 \eeq
Part of our ansatz specifies the orientation of the M2 branes that are smeared inside the M5 worldvolume. We choose the M2s to be oriented along the directions $x^0, x^1, x^2$. Hence, the timelike velocity vector and the three orthonormal spacelike vectors normal to the M2 worldvolume directions inside the M5 branes are
\bea
\label{extrac}
&u^a \pa_a = m^{-1/3} b_0^{-1} (1-{\psi'}^2)^{-1/2} \pa_t
~, ~~
v^a \pa_a = m^{-1/3} b_0^{-1} (\sin\psi)^{-1} \pa_\vartheta
~, \\
&w^a \pa_a = m^{-1/3} b_0^{-1} (\sin\psi \sin\vartheta)^{-1} \pa_\omega
~, ~~
z^a \pa_a = m^{-1/3} b_0^{-1} (\sin\psi \sin\vartheta \sin\omega)^{-1} \pa_\omega
~.
\eea
Inserting this ansatz into the extremal version of the equations \eqref{blackeqsad}-\eqref{blackeqsaf} we obtain the dynamical equations that describe the slow motion of homogeneous polarized anti-M2s in the $\psi$ direction inside the $S^4$ of the background geometry.

A little algebra shows that, when projected along the worldvolume the energy-momentum conservation equations \eqref{blackeqsad} are trivial. The single non-trivial equation in \eqref{blackeqsad} is the $\mu=\psi$ equation, which reads
\beq
\label{extrad}
\cot \psi =  \frac{9}{4 b_0^3} \left( - \tan\theta +  \frac{\sqrt{1 - \psi'^2}}{\cos\theta} \right) - \frac{\psi''}{3 (1 - \psi'^2)}  \frac{1}{\cos^2\theta} 
~.
\eeq

As we noted previously, the current conservation equation \eqref{blackeqsaf} expresses the fact that 
\beq
\label{extrae}
\QQ_5 = \CC r_H^3 \cos\theta
\eeq
is a constant of motion. This equation can be used to eliminate $r_H(t)$ in terms of $\theta(t)$. The second current conservation equation determines the Page charge $\mathbb Q_2$ in terms of the dynamical variables of the problem
\beq
\label{extraf}
\mathbb Q_2 = \frac{27 \pi^2 m}{2} \QQ_5 \left(\frac{1}{3} \cos^3 \psi - \cos \psi + \frac{2}{3} \right) - 2 \pi^2 \CC m b_0^3 r_H^3 \sin \theta \sin^3 \psi
~.
\eeq
Together with \eqref{extrae}, this equation can be used to express $\theta(t)$ in terms of the transverse scalar $\psi(t)$
\beq
\label{extrag}
\tan \theta  =  \frac{1}{b_0^3  \sin^3 \psi} \left[ -\frac{9 p}{\tilde M} + \frac{27 }{4} \left(\frac{1}{3} \cos^3 \psi - \cos \psi + \frac{2}{3} \right)  \right]
~.
\eeq

The equations  \eqref{extrad}, \eqref{extrae}, \eqref{extrag} are a complete set of dynamical equations. It is straightforward to verify that the combination of \eqref{extrad} and \eqref{extrag} follows by Euler-Lagrange variation from the Lagrangian
\beq
\label{extraj}
\LL = \sqrt{1 - \psi'^2}\sqrt{\frac{\hat{H}_0}{96} \sin^6 \psi + \left(\frac{3}{8} f(\psi) - \frac{p}{2 \tilde{M}} \right)^2 } - \frac{3}{8}f(\psi) + \frac{p}{2 \tilde{M}}
~,
\eeq
where
\beq
\label{extrak}
f(\psi) \equiv \frac{1}{3} \cos^3 \psi - \cos \psi +\frac{2}{3}
~.
\eeq
This coincides with the DBI Lagrangian obtained by U-duality from the D4 brane in the probe approximation in \cite{Klebanov:2010qs}. The agreement between the probe DBI expressions and the leading-order extremal blackfold equations that we found here is another example of the supergravity/DBI correspondence.

Our analysis suggests that the use of the DBI equations is a valid long-wavelength approximation in the supergravity regime irrespective of supersymmetry. Therefore, we can use \eqref{extraj} to search for metastable states. In this manner we recover the KP results in the regime of section \ref{validity} directly in supergravity. To find static vacuum configurations we need to extremise the potential that follows from \eqref{extraj}, that is
\beq
\label{extral}
V_{\rm extremal}(\psi) = \sqrt{\frac{\hat{H}_0}{96} \sin^6 \psi + \left(\frac{3}{8} f(\psi) - \frac{p}{2 \tilde{M}} \right)^2 } - \frac{3}{8}f(\psi) + \frac{p}{2 \tilde{M}} 
~.
\eeq

\section{Thermal M-brane anti-blackfolds}
\label{thermal}

From the point of view of holography, temperature can be incorporated in the system at hand in different ways. Let us list here three rough possibilities:
\begin{itemize}

\item One option is to begin by adding temperature to the supersymmetric vacuum of the dual QFT. In the bulk this involves a CGLP black hole with positive M2-brane charge.\footnote{For a construction of smeared black M2-brane solutions that preserve an $SO(5)$ symmetry see \cite{Dias:2017opt}.} Then, one can analyse the existence and properties of a metastable state in this thermal environment. In the bulk, an analysis based on the probe approximation would entail at leading order the use of a DBI-type action for a wrapped M5 brane in the background of the CGLP black hole. 

\item A second option that focuses more directly on thermal effects on the metastable state itself goes along the following lines. In the bulk description, we can either consider solutions in the probe approximation using a thermalised DBI-like (or PST-like) effective action in CGLP, or in the supergravity regime we can attempt to construct a wrapped M5 black hole with negative M2 charge that asymptotes to the supersymmetric background. In this section, we will focus on the second approach using blackfold techniques. The fundamental difference between this bullet point and the previous one is that as one turns off the anti-brane charge, in the first case one recovers a thermal state of the dual QFT, whereas in the second one recovers a supersymmetric ground state of the dual QFT.

\item A third, more general, option is to thermalise all the sectors of the system at the same time.%
\footnote{This approach was taken for example in \cite{Grignani:2010xm} which considers the thermalized version of the
BIon solution by analyzing a D3-F1 blackfold in hot flat space  and in \cite{Grignani:2012iw}
which considers  an F1 blackfold in the AdS black hole background to study finite temperature Wilson loops.}
This would entail in the bulk the construction of a black hole solution that describes the backreaction of a thermally excited wrapped M5 brane in the background of the CGLP black hole. One could also try to capture aspects of this case with blackfold techniques, but we will not explore this possibility in this case. 
\end{itemize}

Adding temperature to the effective actions of weakly coupled open strings is a notoriously difficult problem that involves open string loop computations (we refer the reader to \cite{Grignani:2013ewa} for a relevant discussion). In that sense, implementing the option of the second bullet point with a DBI-like probe analysis is not a straightforward exercise. In the supergravity regime, however, the blackfold equations allow the incorporation of thermal effects rather easily. In this section we present the explicit form of the non-extremal blackfold equations and discuss ways to obtain static thermal vacua from the extremisation of thermal effective potentials.

\subsection{Non-extremal equations}
\label{theqs}

For concreteness, let us consider again a time-dependent ansatz for a wrapped M5. To obtain the thermal version of the equations in section \ref{sec:extremal} we need to find the explicit form of the equations \eqref{blackeqsad}-\eqref{blackeqsaf} without implementing the extremal limit $r_0\to 0$, $\alpha\to \infty$ with $r_H$ fixed. In this case, $r_0(t)$ and $\alpha(t)$ are independent dynamical variables.

The current conservation equation \eqref{extrag} remains the same 
\beq
\label{theqsaaa}
\tan \theta  =  \frac{1}{b_0^3  \sin^3 \psi} \left[ -\frac{9 p}{\tilde M} + \frac{27 }{4} \left(\frac{1}{3} \cos^3 \psi - \cos \psi + \frac{2}{3} \right)  \right]
~.
\eeq
Eq.\ \eqref{extrae} becomes
\beq
\label{theqsaa}
\QQ_5 = \CC r_0^3 \cos\theta \sinh\alpha \cosh\alpha
\eeq
and can be used to eliminate $r_0$ in terms of $\alpha$ and $\theta$. The non-extremal analog of eq.\ \eqref{extrad} is
\begin{equation}
\begin{split}
\label{theqsab}
&\frac{\psi''}{1 - \psi'^2}\left( \frac{4}{9} + \frac{1}{3} \sinh^2 \alpha \right) + \cot \psi \left( \frac{1}{3} + \cos^2 \theta \sinh^2 \alpha \right) 
\\
&=  \frac{9}{4 b_0^3} \cos\theta \sinh\alpha \left( - \sin \theta \sinh \alpha +  \sqrt{1 - \psi'^2}  \cosh \alpha \right)
~.
\end{split}
\end{equation}

These are three equations for four unknowns. Unlike the extremal case where these equations follow from a variational principle, in the non-extremal case there is no obvious candidate of an effective action for arbitrary time-dependent configurations.

In what follows, we focus on static configurations where the equation \eqref{theqsab} simplifies to
\beq
\label{theqsac}
\cot \psi \left( \frac{1}{3} + \cos^2 \theta \sinh^2 \alpha \right) 
=  \frac{9}{4 b_0^3} \cos\theta \sinh\alpha \left( - \sin \theta \sinh \alpha +   \cosh \alpha \right)
~.
\eeq
The solutions of eqs.\ \eqref{theqsaaa}, \eqref{theqsac} at fixed $p/\tilde M$ are parametrised by a free constant. This could be a non-extremality parameter like $r_0$ or $\alpha$, or a more physically motivated thermodynamic parameter like the total entropy $S$ or the global temperature $T$.

The solutions of \eqref{theqsaaa}, \eqref{theqsac}, and their properties, will be discussed in detail in section \ref{meta}. In the rest of this section we explain how to obtain these solutions as extrema of suitable effective potentials. A different potential is formulated for each parameter that we choose to keep fixed. We will discuss three kinds of potentials: $V_T$ where the global temperature $T$ is kept fixed, $V_{S}$ where the total entropy $S$ is kept fixed and $V_{\alpha}$ where the parameter $\alpha$ is kept fixed. The first two thermodynamic potentials have appeared before (see e.g.\ \cite{Emparan:2011hg,Armas:2018ibg}). The third one, which is new, is non-thermodynamic.

\subsection{Interlude on effective potentials}
\label{potentials}

\subsubsection{Comments on effective thermodynamic potentials}

Before we present specific effective potentials for the M2-M5 configurations of interest, it is useful to first comment on a slightly more general problem. The general blackfold equations describe an effective fluid on a dynamical hypersurface. Let us assume that we are interested in stationary solutions of these equations. For such solutions there is a worldvolume Killing vector field ${\boldsymbol k}^a$, which is assumed to be the pullback of a background Killing vector field ${\boldsymbol k}^\mu$. It has been argued in \cite{Emparan:2009at,Emparan:2011hg,Armas:2018ibg} that by using standard thermodynamic quantities it is possible to formulate effective actions of the transverse scalars whose extrema reproduce the profiles of stationary solutions. In these actions the intrinsic degrees of freedom of the effective fluid are integrated out and the variational problem is restricted to stationary configurations. These actions are guaranteed to produce correct stationary solutions if they recover the currents of the fluid under general variations of the background fields.

For concreteness, let us focus on the case of interest: M2-M5 blackfolds on the CGLP background. We can obtain an effective action by varying over stationary configurations at a fixed global temperature $T$ in the following manner. By definition, the global temperature is related to the local temperature $\TT$ of the effective fluid in \eqref{forceaea} as $T= |{\boldsymbol k}| \TT$. For the Killing vector ${\boldsymbol k}^a$ we have ${\boldsymbol k}^a \pa_a = m^{1/3}b_0 \, \partial_t$ and $u^a = {\boldsymbol k}^a/|{\boldsymbol k}|$ ---hence, $|{\boldsymbol k}|=m^{1/3}b_0$. Variations of the background lead to variations of the induced metric $\delta \gamma_{ab}$, the pulled-back three-form gauge potential $\delta A_{3abc}$ and its dual $\delta A_{6abcdef}$, defined such that $G_7=dA_6+A_3\wedge G_4/2$. These variations are performed keeping $\mathbb Q_2$, $\QQ_5$ and $T$ fixed. Let us denote $\gamma^\perp_{ab} \equiv v_a v_b + w_a w_b + z_a z_b$ the projector onto worldvolume directions perpendicular to the dissolved M2 directions. Keeping $\mathbb Q_2$, $T$ and $\QQ_5 T^3$ fixed under variations imply the variational properties
\beq
\begin{split}
\label{thermoaa}
&\delta \tan\theta = (v \wedge w \wedge z)^{abc} \delta A_{3abc} - \frac{1}{2} \tan\theta \, \gamma^{\perp ab} \delta \gamma_{ab}~~, \\
&\delta r_0 = -\frac{1}{2} r_0 u^a u^b \delta \gamma_{ab} - r_0 \sinh\alpha \cosh\alpha \, \delta(\tanh\alpha)~~, \\
&\delta (\tanh\alpha) = \frac{3}{2} \frac{\tanh \alpha}{1-\sinh^2\alpha} u^a u^b \delta \gamma_{ab}
+ \frac{\tanh \alpha}{1-\sinh^2\alpha} \sin\theta \cos\theta\, \delta (\tan \theta)~~,
\end{split}
\eeq
respectively. Under such variations one can easily show that the thermodynamic effective action 
\beq
\label{thermoad}
\begin{split}
{\mathcal S}_T =& - \int_{\MM_6} d^6 \sigma \sqrt{-\gamma} \, \FF + \QQ_5 \int_{\MM_6} \mathbb{P}[A_6]+\mathbb Q_2\int_{\MM_3} \mathbb{P}^{||}[A_3]~~,\\
=& - \int_{\MM_6} d^6 \sigma \sqrt{-\gamma} \, \FF + \QQ_5 \int_{\MM_6}\left(\mathbb{P}[A_6]+\frac{1}{2}\mathbb{P}[A_3\wedge A_3]+\frac{\Phi_2}{\Phi_5}dV_\perp\wedge\mathbb{P}^{||}[A_3]\right)~~,
\end{split}
\eeq
reproduces the correct currents
\beq
\label{thermoae}
\delta {\mathcal S}_T = \int_{\MM_6} d^6 \sigma \sqrt{-\gamma} \left( \frac{1}{2} T^{ab}\delta \gamma_{ab} + \tilde J_3^{abc}\delta A_{3abc} + \JJ_6^{a_1\cdots a_6} \delta A_{6a_1\cdots a_6} \right)
~.
\eeq
In \eqref{thermoad} $\FF=\varepsilon- \TT s$ is the free energy \eqref{forceaed}, $\mathbb{P}[A_6]$ is the pullback of the background six-form $A_6$, $\MM_6$ is the six-dimensional worldvolume of the effective theory, $\mathbb{P}^{||}[A_3]$ is the pullback of the background $A_3$ onto $\mathcal{M}_3$ and $dV_\perp=\sqrt{\gamma_\perp}dv\wedge d w\wedge dz$ is the volume form on $\mathcal{M}_3^\perp$. 

Invariance of \eqref{thermoad} under gauge transformations $\delta A_3=d\Lambda_2$ and $\delta A_6=d\Lambda_5$ for gauge parameters $\Lambda_2,\Lambda_5$ leads to the conservation equations for $ \tilde J_3$ and $\JJ_6$ respectively as in \eqref{blackeqsae}-\eqref{blackeqsaf}. The last term in \eqref{thermoad} vanishes for the specific configurations that we are interested in, since $\mathbb{P}^{||}[A_3]=0$. However, in order to extract the correct currents via a variational principle, it is required.

In \eqref{thermoad} it is implicitly assumed that we have implemented all constraints from the constant $\mathbb Q_2$, $\QQ_5$ and $T$ together with a stationary ansatz for the vectors $u^a, v^a, w^a, z^a$ and that we have expressed $r_0$, $\alpha$, $\theta$ in terms of the transverse scalars. The resulting action is an action of the transverse scalars alone.\footnote{It is also possible to write an action for M2-M5 branes that does not make assumptions about the background or how the M2 branes are embedded into the M5. In this case, additional dynamical fields must be introduced as in \cite{Armas:2018atq, Armas:2018zbe}.} By varying it with respect to the transverse scalars we are guaranteed to obtain equations that lead to the correct stationary solutions of the blackfold equations. Explicit formulae for wrapped M5 branes will appear in the next subsection. 

The effective action \eqref{thermoad} has a well defined extremal limit $T\to0$, reducing to the PST action \cite{Pasti:1996vs, Pasti:1997gx} (multiplied by the number of M5-branes  $N_5$) when all worldvolume gauge fields have been integrated out. However, the existence of a maximum temperature \eqref{forceaea} and, in cases of bound states with a Hagedorn temperature such as the D3-NS5 brane \cite{Armas:2018rsy} for which $T=0$ does not describe all extremal solutions, the potential at fixed $T$ is unsuitable for describing the entire phase space of off-shell configurations. Instead, defining $\BB_5$ as the spatial part of the worldvolume $\MM_6$, a closely related effective action, where we keep the total entropy 
\beq
\label{thermoaf}
S =  \int_{\BB_5} \sqrt{-\gamma} \, s \, u^t
\eeq 
fixed, is more appropriate and can be obtained by Legendre transforming \eqref{thermoad} yielding
\beq
\label{thermoag}
{\mathcal S}_S = -\int_{\MM_6} d^6 \sigma \sqrt{-\gamma} \, \varepsilon + \QQ_5 \int_{\MM_6} \mathbb{P}[A_6]+\mathbb Q_2\int_{\MM_3} \mathbb{P}^{||}[A_3]
~.
\eeq 
The Wick rotated version of \eqref{thermoag} corresponds to the total energy in the system as we explicitly show in App.~\ref{smarr}. Different choices of thermodynamic ensembles, where other global chemical potentials are kept fixed, are also possible and we have deferred this analysis to  App.~\ref{smarr}.

\subsubsection{Potential at fixed temperature}
\label{fixedT}

In this subsection we present the precise form of the potential $V_T$ for M5 branes wrapping $S^3$ in \eqref{setupae}. The black M5s of interest are characterised by the local temperature $\TT =\frac{3}{4\pi r_0 \cosh\alpha}$. Eliminating $r_0$ with the use of eq.\ \eqref{theqsaa} we can write
\beq
\label{potTaa}
\TT^3 = \frac{27 \CC}{64\pi^3 \QQ_5} \T^3~, ~~~
\T^3 \equiv \frac{\cos\theta \sinh\alpha}{\cosh^2\alpha}
~.
\eeq
We will use $\T$ to express all the relevant formulae. Dividing ${\mathcal S}_T$ by the infinite volume of the $\IR^{3,1}$ part of the M5 worldvolume and an overall constant factor of $36 \pi^2 m^2 b_0^3 \QQ_5$ we obtain the potential
\beq
\label{potTab}
V_T(\psi) = \frac{b_0^3 \sin^3\psi (1 + 3 \sinh^2 \alpha(\psi) )}{54 \cos\theta(\psi) \sinh\alpha(\psi) \cosh\alpha(\psi) } - \frac{3}{8} f(\psi) 
~.
\eeq
In this formula $\theta(\psi)$ is obtained by using eq.\ \eqref{theqsaaa}. $\alpha(\psi)$ is determined by combining \eqref{potTaa} and \eqref{theqsaaa}. The function $f(\psi)$ was defined in eq.\ \eqref{extrak}. The potential $V_T$ depends parametrically on $p/\tilde M$ and $\T$. One can show by direct evaluation that the equation $\frac{dV_T}{d\psi}=0$ is equivalent to the equation that follows from the blackfold equations \eqref{theqsaaa}, \eqref{theqsaa}, \eqref{theqsac}. In particular, when $\T=0$ we recover the extremal vacua of Klebanov and Pufu.

There are two intricacies of the fixed-$T$ potential that are worth highlighting. The first one is that eq.\ \eqref{potTaa} has in general two solutions of $\alpha$ for a given angle $\psi$ at a fixed value of $\T$. 
The two solutions are
\beq
\label{potTac}
\tanh\alpha_\pm (\psi) = \sqrt{\frac{1}{2} \pm \sqrt{\frac{1}{4} - \frac{\T^6}{\cos^2\theta(\psi)}}}
~.
\eeq
The branch of $\alpha_+$ is valid for $\alpha \geq \alpha_*$ and the branch of $\alpha_-$ for $\alpha\leq \alpha_*$. The critical value $\alpha_*$ defines a point where $\alpha_+ = \alpha_-$, i.e.\ a point where $\cos^2\theta(\psi) = 4 \T^6$. Numerically, $\alpha_* \simeq 0.881374$. Notice that these solutions are real only when 
\beq
\label{potTad}
|\cos\theta(\psi)| \geq 2 \T^3
~.
\eeq
This inequality imposes a constraint on the domain of $\psi$ where the potential $V_T(\psi)$ in \eqref{potTab} can be defined sensibly.

\begin{figure}[t!]
\begin{center}
\includegraphics[width=10cm]{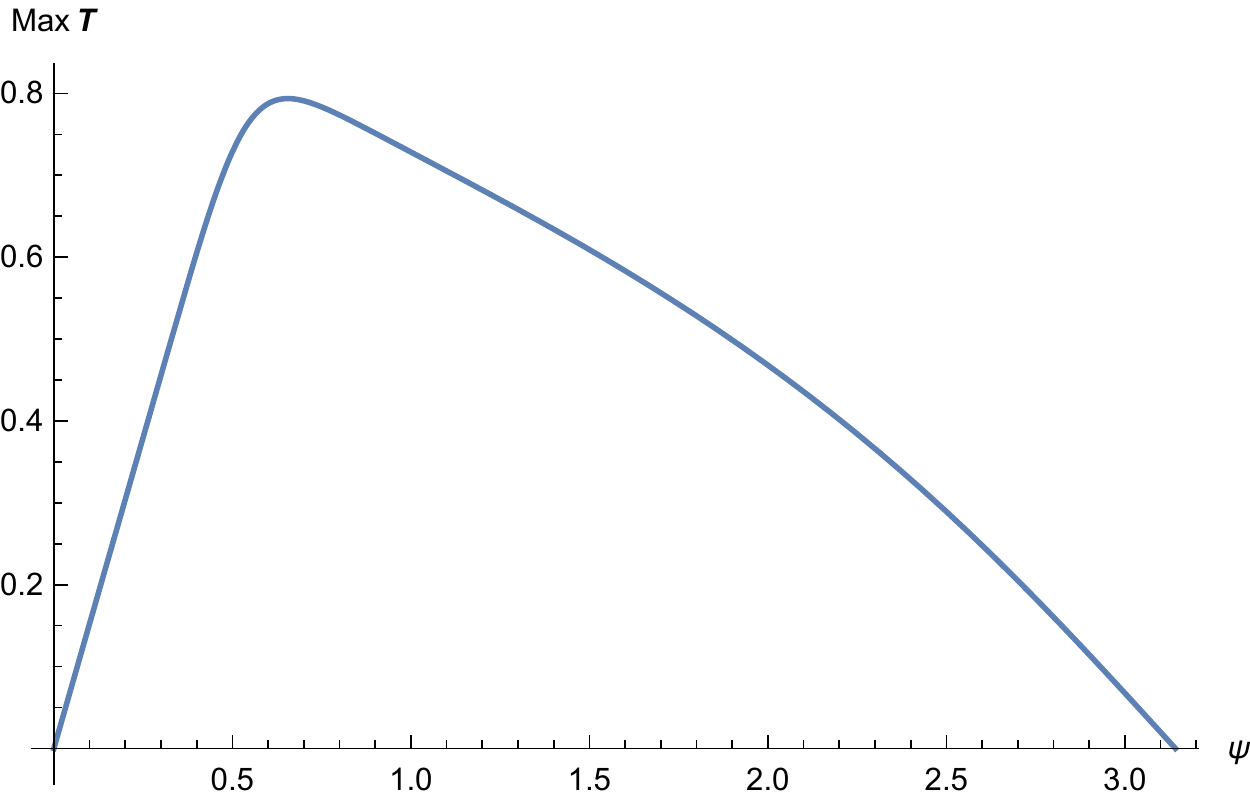}
\end{center}
\caption{A plot of the maximum possible value of the temperature $\T$ of the M2-M5 branes at each angle $\psi$ for $p/\tilde{M} = 0.03$.}
\label{Fig_maxtemp}
\end{figure}

The second related feature is that the temperature of the wrapped M5s at a given angle $\psi$ has a maximum possible value. This follows immediately from \eqref{potTaa} and the fact that the function $\sinh\alpha/\cosh^2\alpha$ has a maximum value of $1/2$. A plot of the maximum temperature at a given angle $\psi$ appears in Fig.\ \ref{Fig_maxtemp}.

\subsubsection{Potential at fixed entropy}

The action \eqref{thermoag} allows us to formulate a potential whose extrema determine the equilibria of the wrapped M5 branes at a fixed total entropy. In the case at hand the total entropy \eqref{thermoaf} is given by the expression
\beq
\label{potSaa}
S = \frac{8\pi^3 m^{\frac{5}{3}} b_0^5 \QQ_5^{\frac{4}{3}}}{3\, \CC^{\frac{1}{3}}} \bS
~,~~
\bS^3 \equiv \frac{\sin^9\psi}{\cos^4 \theta \sinh^4\alpha \cosh\alpha}
~.
\eeq
We express all relevant quantities using the properly normalised entropy $\bS$. Dividing ${\mathcal S}_S$ in \eqref{thermoag} by the infinite volume of the $\IR^{3,1}$ part of the M5 worldvolume and the overall factor $36 \pi^2 m^2 b_0^3 \QQ_5$ we obtain the potential
\beq
\label{potSab}
V_S (\psi) = \frac{b_0^3 \sin^3\psi (4 + 3 \sinh^2 \alpha(\psi) )}{54 \cos\theta(\psi) \sinh\alpha(\psi) \cosh\alpha(\psi) } - \frac{3}{8} f(\psi) 
~.
\eeq
Again, one can verify by direct computation that the extrema of this potential reproduce the correct static solutions of the blackfold equations at fixed total entropy $S$. The potential $V_S$ depends parametrically on $p/\tilde M$ and the entropy $\bS$. At $\bS=0$ the potential \eqref{potSab} reduces to the potential that follows from the DBI action.

In this case the potential is well defined in the whole range of angles $\psi$. We will present numerical plots of the potential in different regimes of parameters in the next section. The same type of fixed-entropy potential was computed for the wrapped NS5 branes in the Klebanov-Strassler background of type IIB string theory in \cite{Armas:2018rsy}.

\subsubsection{Potential at fixed $\alpha$}

The above discussion demonstrates that one can consider effective potentials in different ensembles. All of them reproduce the same static configurations as the original blackfold equations but the off-shell shape of the potential in each case is different. It is natural to ask whether it is possible to define a potential that keeps some other quantity constant, possibly one that does not have a straightforward thermodynamic interpretation. When we solve the combination of eqs.\ \eqref{theqsaaa}, \eqref{theqsac}, technically the most convenient choice would be to solve them at a fixed value of $\alpha$. $\alpha=\infty$ would be the extremal case and $\alpha=0$ the exact opposite.

One can show by direct computation that the following fixed-$\alpha$ potential does the job:
\beq
\label{potAaa}
V_\alpha (\psi) = \frac{1}{18} b_0^3 \sin^3 \psi \frac{1}{\cos \theta(\psi)} - \frac{3}{8} \coth \alpha\, f(\psi) + \frac{1}{\sinh^2 \alpha} H (\psi)
\eeq
with
\beq
\label{potAab}
H(\psi) = \int_{\psi_0}^\psi d \chi \left( \cot \chi \sqrt{ \frac{\hat{H}_0}{96} \sin^6 \chi + \left(\frac{3}{8} f(\chi) - \frac{p}{2 \tilde{M}} \right)^2 } \right)
~.
\eeq
The constant $\psi_0$ in the lower limit of the integration in \eqref{potAab} is arbitrary. Its value determines an arbitrary additive constant to the potential. As before, we obtain $\theta(\psi)$ by solving the eq.\ \eqref{theqsaaa}. As a trivial check, notice that $V_\alpha$ reduces to the extremal potential \eqref{extral} when $\alpha \to \infty$ (in that case the last term in $V_\alpha$ vanishes).

By varying the values of $\alpha$ we obtain the full range of static wrapped M5-brane configurations that we would obtain directly from the blackfold equations. The same overall set of static configurations can be obtained by extremising either of the potentials $V_T$ and $V_S$ for different values of $T$ and $S$. In that sense, all the potentials that we described above are equivalent. 

The off-shell shape of each potential is different. One should exercise some caution when employing the full shape of the potential to make statements about, say, the stability of the different vacua. Since the entropy current is conserved in our leading order ideal hydrodynamic effective theories, time-dependent solutions will naturally evolve conserving the total entropy. This suggests that the off-shell shape of the fixed-$S$ potential contains correct information about the stability of the vacua we find (at least within the homogeneous ansatz of wrapped M5s that we used). In the next section, we plot the fixed-$\alpha$ potential and show that it shares the same qualitative features as the fixed-$S$ potential.

\section{M-brane metastability at finite temperature}
\label{meta}

We are now in position to determine in detail what happens to the KP vacua once we turn on the temperature. We will discuss the non-extremal physics from the perspective of all the potentials presented in the previous section.

\begin{figure}[t!]
\begin{center}
\includegraphics[width=12cm]{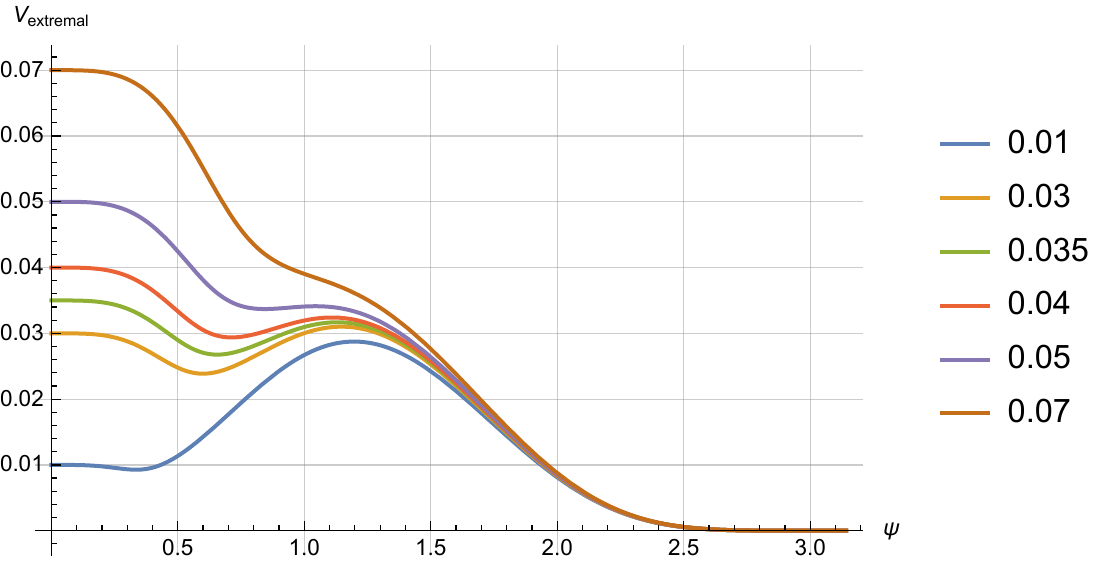}
\end{center}
\caption{A plot of the extremal potential $V_{\rm extremal}$ in \eqref{extral} as a function of $\psi$. Different colors depict the plot for different values of the $p/\tilde M$ (the values of $p/\tilde M$ for each color are quoted in the legend on the right).}
\label{extremal}
\end{figure}

\subsection{Vacua and transitions}
\label{metavacua}

For reference, Fig.\ \ref{extremal} depicts the extremal potential, first obtained in \cite{Klebanov:2010qs}. There is a clearly visible metastable vacuum for $p/\tilde M \leq \mathfrak p_* \simeq 0.0538$. In this regime there are also two unstable extrema: one at $\psi=0$ and another in the vicinity of $\psi\simeq 1.2$. The point $\psi=0$ is outside the regime of validity of our long-wavelength approximations.

\begin{figure}[t!]
\begin{center}
\includegraphics[width=10cm]{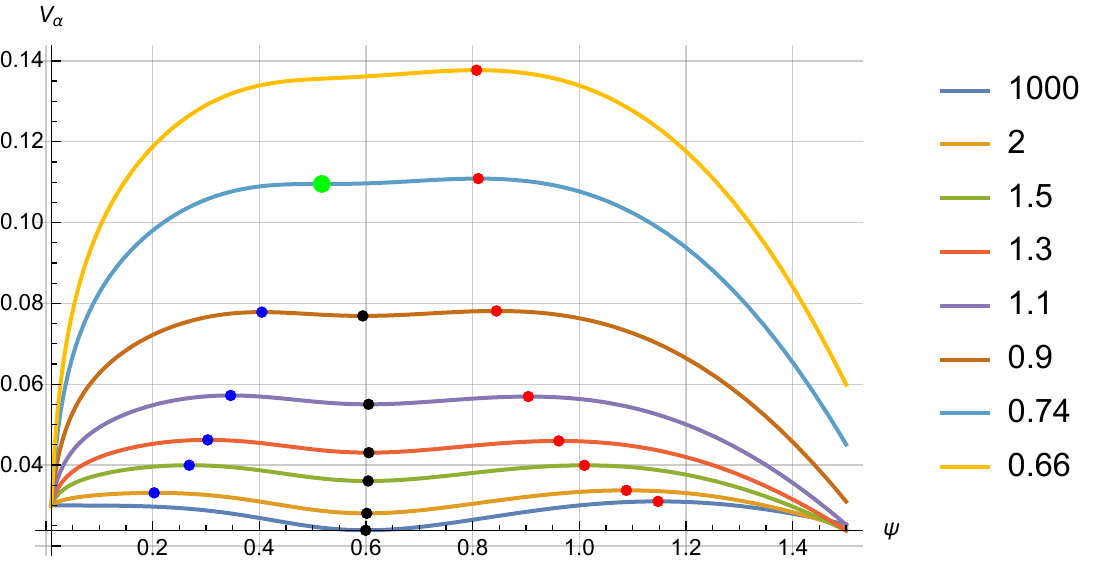}
\end{center}
\caption{Plots of the non-extremal potential $V_\alpha$ as a function of $\psi$ for $p/\tilde M =0.03$. The range of the plot is restricted in the region $\psi\in (0,1.5)$ where the most interesting physics occurs. Different colors depict the potential at different values of the non-extremality parameter $\alpha$ (the specifics of these values are listed in the legend). The blue dots indicate the unstable fat M5 vacua near the north pole $(\psi=0)$. The black dots indicate the metastable vacuum. The red dots indicate a second unstable vacuum (thin M5 branch). The green dot at $\alpha \simeq 0.7424$ is a merger point of the blue and black vacua.}
\label{Valpha_3}
\end{figure}

\begin{figure}[t!]
\begin{center}
\includegraphics[width=7.6cm]{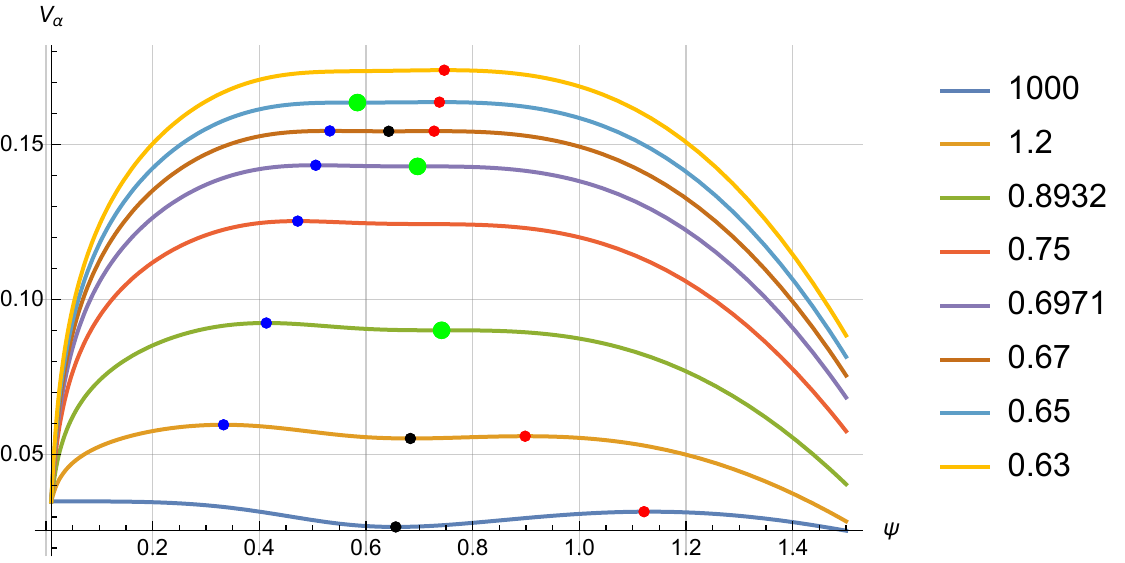}
\hspace{0.1cm}
\includegraphics[width=7.6cm]{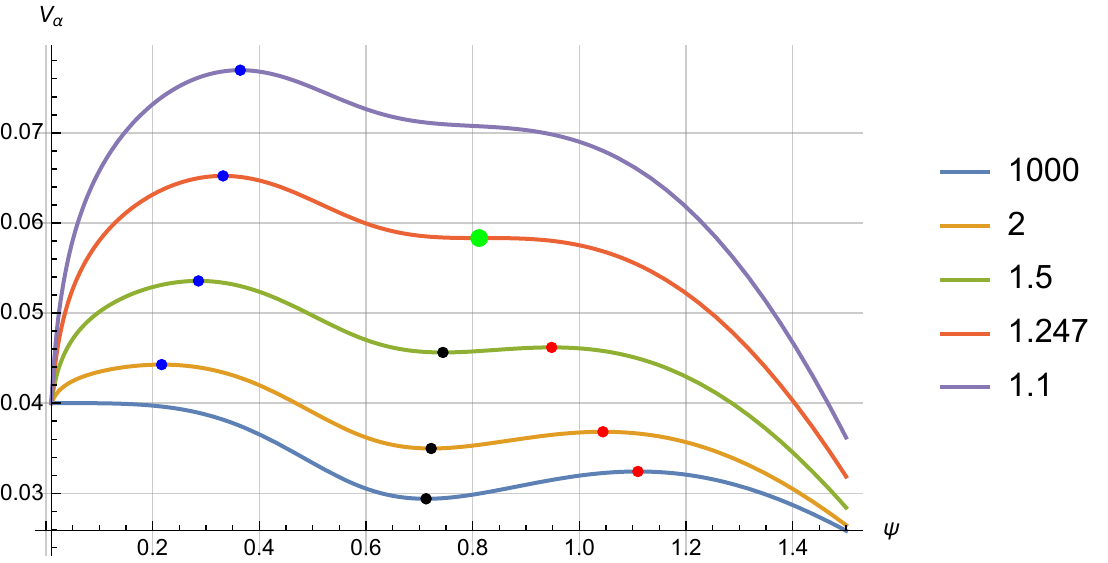}
\end{center}
\caption{On the left, we plot the non-extremal potential $V_\alpha$ as a function of $\psi$ for $p/\tilde M =0.035$. Different colors depict the potential at different values of the non-extremality parameter $\alpha$. Once again, the blue dots indicate an unstable vacuum near the north pole $(\psi=0)$ (fat M5 branch) and the black dots the metastable vacuum. The red dots indicate a second unstable vacuum (thin M5 branch). In this case there are three green dots. At $\alpha \simeq 0.8932$ and $\alpha \simeq 0.6971$ they represent merger points of a metastable state with a red unstable thin M5 state. At $\alpha \simeq 0.65$ the green dot represents a merger with a blue unstable fat M5 state. On the right, we plot the non-extremal potential $V_\alpha$ as a function of $\psi$ for $p/\tilde M =0.04$. The plotted values of $\alpha$ are listed in the legend. In this regime there is a single green dot at $\alpha \simeq 1.247$, which is a merger point of the black metastable state with the red unstable thin M5 state.}
\label{Valpha_35_4}
\end{figure}

In what follows we focus on the `metastable regime' $p/\tilde M\in (0,\mathfrak p_*)$ and examine how thermal effects modify the stable and unstable vacua. It is technically convenient to start with the analysis of the blackfold equations at fixed $\alpha$, where the plots of the potential $V_\alpha$ (in \eqref{potAaa}, \eqref{potAab}) exhibit the extrema most clearly. In Figs.\ \ref{Valpha_3} and \ref{Valpha_35_4},  we present plots of $V_\alpha$ at three different values of $p/\tilde M$: 0.03, 0.035 and 0.04. Curves with different colors represent the form of the potential at the same $p/\tilde M$ for different values of $\alpha$. The dots indicate the location of the extrema and the color of the dots the nature of the solution at those extrema. We use the following conventions:
\begin{itemize}
\item A blue dot represents an unstable solution in the vicinity of the north pole at $\psi=0$. This is a black M5 brane solution wrapping an $S^3$ with a small radius compared to the Schwarzschild radius (further details on this aspect will appear in the next subsection). We call this type of solutions {\it fat M5 branes}.  
\item A black dot represents a metastable solution. These solutions are thermalised versions of the KP metastable state. 
\item A red dot represents an unstable wrapped M5 black brane whose $S^3$ radius is large compared to the Schwarzschild radius. We call this type of solutions {\it thin M5 branes}.
\item A green dot represents the merger of an unstable state with a metastable state.
\end{itemize}

Depending on the regime of $p/\tilde M$ the system exhibits three different types of bifurcations. 

\vspace{0.2cm}
\noindent
{\bf Regime I: small $p/\tilde M$.}
The first type occurs for $p/\tilde M \in (0,\mathfrak p_1)$. Numerically, we have determined $\mathfrak p_1 \simeq 0.0345$. The characteristic behaviour of this regime appears in Fig.\ \ref{Valpha_3}. The bottom blue curve is a near-extremal curve at $\alpha=1000$. As we decrease $\alpha$ (and therefore increase the non-extremal effects) we observe the gradual convergence of the fat unstable branch towards the metastable branch. They merge at a small value of $\alpha$ ($\alpha\simeq 0.7424$ in the case of Fig.\ \ref{Valpha_3}) at $\psi \simeq 0.5$ which corresponds to the renormalised temperature $\T \simeq 0.73873$. At even smaller values of $\alpha$ only the unstable thin M5 brane branch (red dot) remains. In this regime we observe the same saddle-node type bifurcation that was observed in the case of polarised anti-D3 branes in the Klebanov-Strassler background \cite{Armas:2018rsy}. In the next subsection we will present quantitative evidence that suggests that this type of merger is driven by properties of the horizon geometry.

\vspace{0.2cm}
\noindent
{\bf Regime II: intermediate $p/\tilde M$.}
Interestingly, unlike the polarised anti-D3s in Klebanov-Strassler, in the M-theory case at hand there are two additional types of transitions that point towards qualitatively different properties of the dual three-dimensional QFT. A more involved transition pattern occurs for $p/\tilde M \in (\mathfrak p_1, \mathfrak p_2)$. Numerically, we obtain $\mathfrak p_2 \simeq 0.0372$. This is a small window where, as we decrease $\alpha$, {\it three} consecutive saddle-node-type bifurcations occur. First, the metastable state merges with the red thin black M5 state on the right of the plot (in Fig.\ \ref{Valpha_35_4} on the left this occurs at $\alpha \simeq 0.8932$, $\psi\simeq 0.74173$ and $\T \simeq 0.78917$). This type of transition is qualitatively different compared to the transition in Fig.\ \ref{Valpha_3}. It bears a strong resemblance to the zero temperature transition in Fig.\ \ref{extremal} when $p/\tilde M$ crosses the threshold for the existence of a metastable vacuum. For a range of lower values of $\alpha$ only the fat unstable M5 brane state (blue dots) exists. Then, at another saddle-node-type bifurcation a metastable state and a thin unstable state re-appear out of nothing (in Fig.\ \ref{Valpha_35_4} on the left this occurs at $\alpha\simeq 0.6971$, $\psi \simeq 0.69662$ and $\T\simeq 0.78318$). Subsequently, at even lower values of $\alpha$ the new metastable state starts moving closer to the fat unstable state. A third and final merger between the fat and metastable states occurs, which is qualitatively of the same character as in regime I. In Fig.\ \ref{Valpha_35_4} on the left this merger occurs at $\alpha\simeq 0.65$, $\psi\simeq 0.58389$ and $\T\simeq 0.75475$.

\vspace{0.2cm}
\noindent
{\bf Regime III: high $p/\tilde M$.}
There is a third regime, where $p/\tilde M\in (\mathfrak p_2, \mathfrak p_*)$. In this case, there is only one merger, which is a merger between the metastable state and the red thin unstable state. In Fig.\ \ref{Valpha_35_4} on the right this merger occurs at $\alpha\simeq 1.247$ at $\psi\simeq 0.81220$, $\T\simeq 0.75445$. As we noted above, this type of thermal transition does not occur in the case of polarised anti-D3 branes in the Klebanov-Strassler background. In the next subsection we present evidence suggesting that properties of the horizon geometry play a less important role in this type of merger.

\begin{figure}[t!]
\begin{center}
\includegraphics[width=7.3cm]{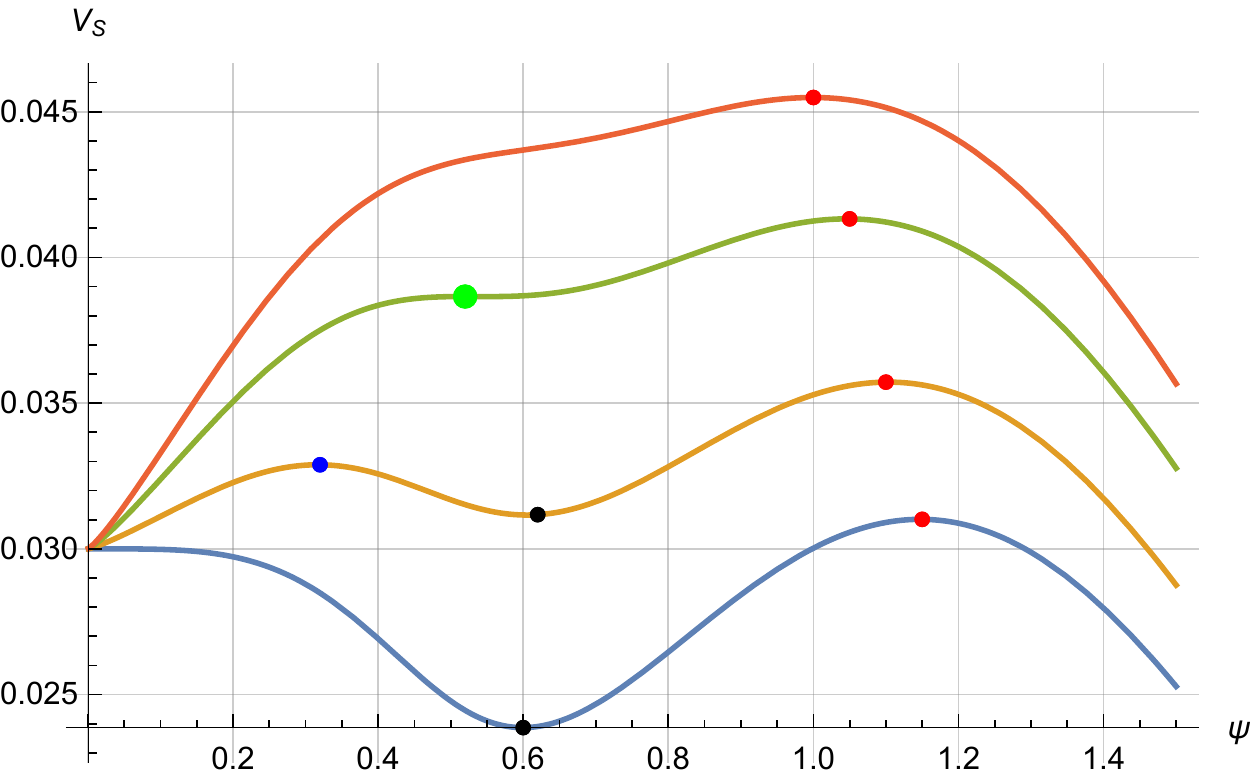}
\hspace{0.2cm}
\includegraphics[width=7.3cm]{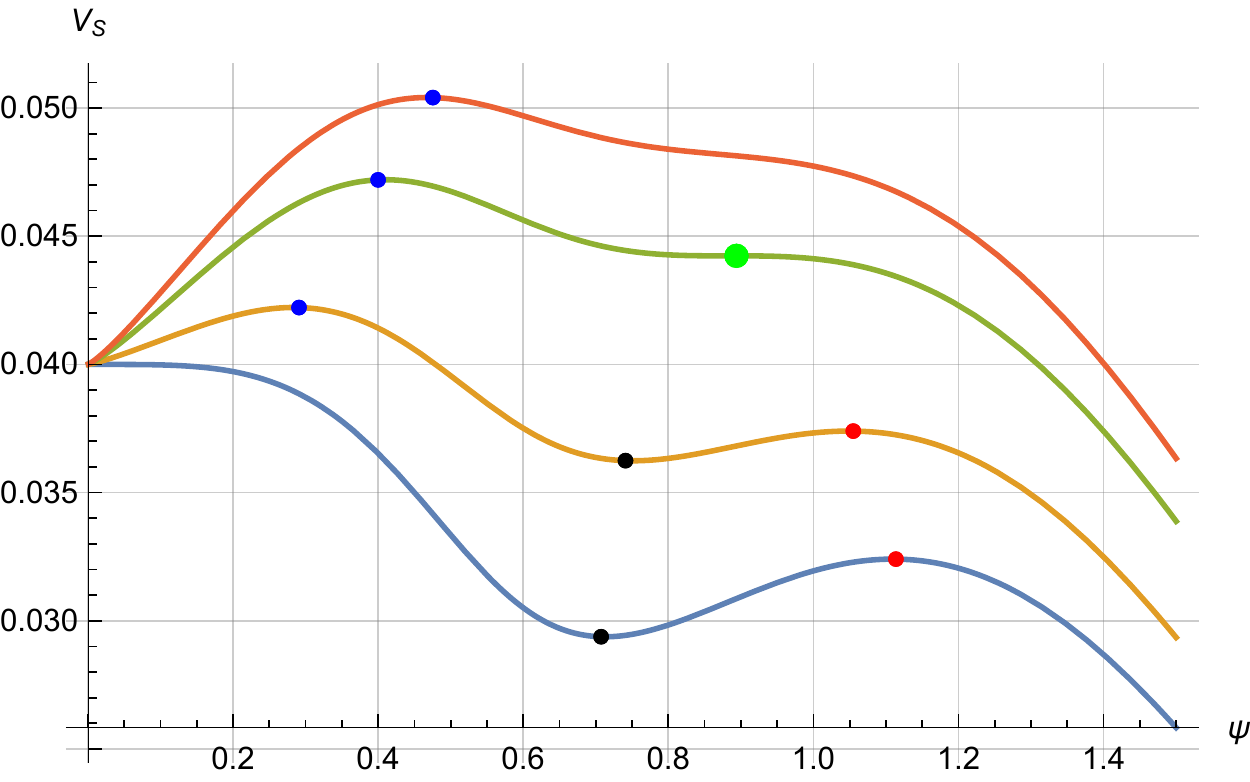}
\end{center}
\caption{Plots of the fixed-$S$ counterparts of the plots in Figs.\ \ref{Valpha_3} and \ref{Valpha_35_4} on the right. The left plot depicts $V_S$ at $p/\tilde M =0.03$ and the right plot $V_S$ at $p/\tilde M=0.04$.}
\label{Ventro}
\end{figure}

The corresponding analysis of the system at fixed total entropy $S$ with the use of the potential $V_S$ reveals exactly the same qualitative and quantitative features. In Fig.\ \ref{Ventro} we present the fixed-$S$ counterparts of the plots in Figs.\ \ref{Valpha_3} and \ref{Valpha_35_4} on the right. It is visually harder to observe the three transitions in the plot of $V_S$ at $p/\tilde M = 0.035$, so we did not include this value in Fig. \ref{Ventro}. As we noted previously, the $V_S$ potential is better motivated physically compared to the $V_\alpha$ potential.

\begin{figure}[t!]
\begin{center}
\includegraphics[width=7.6cm]{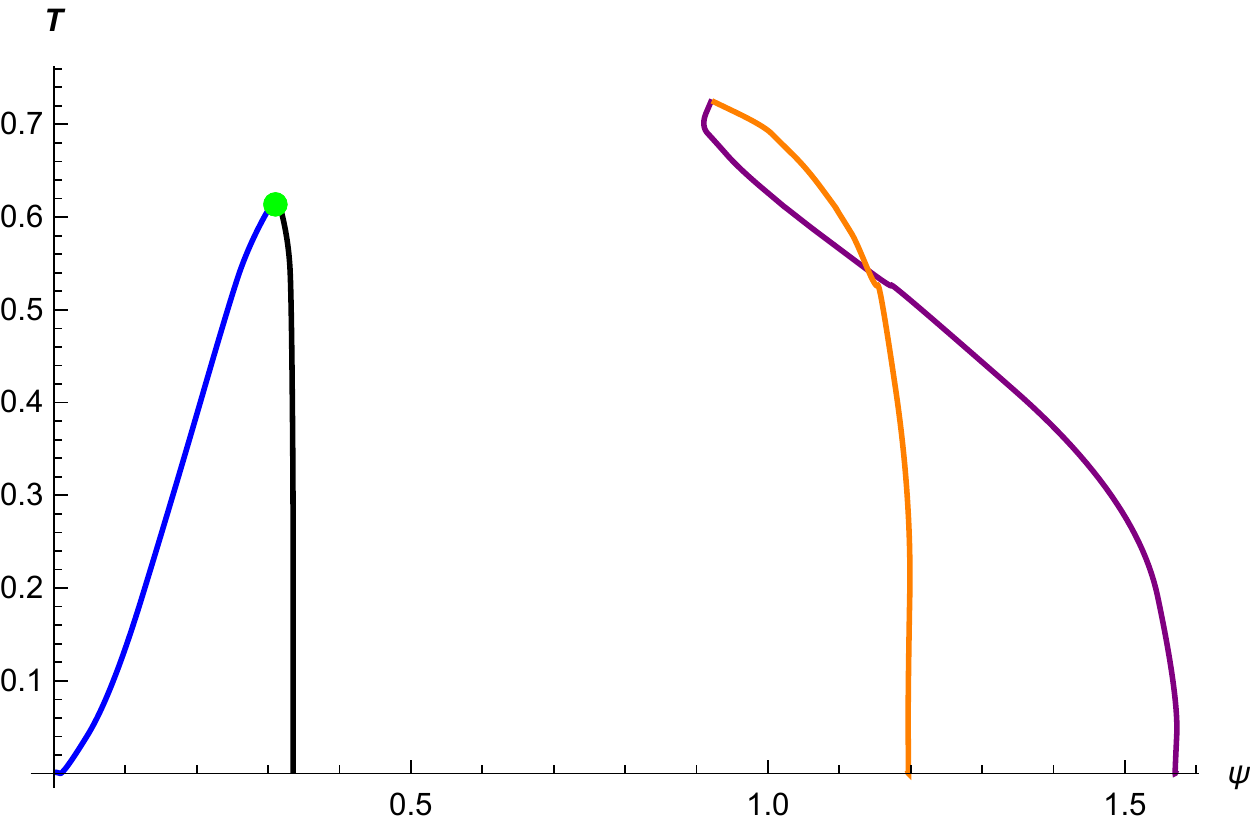}
\hspace{0.1cm}
\includegraphics[width=7.6cm]{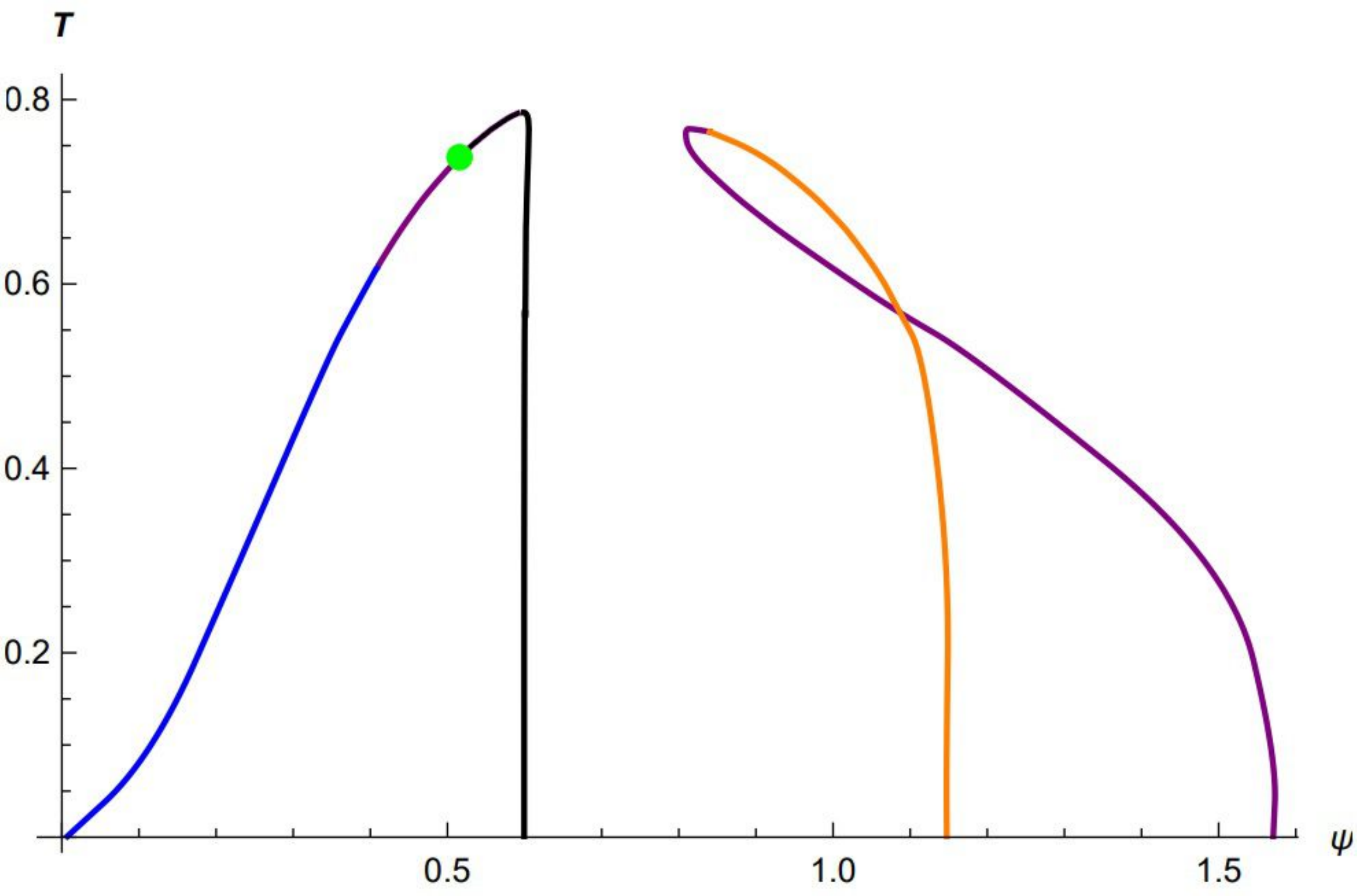}
\end{center}
\caption{On the left, we plot the temperature of all the static M5-brane configurations as a function of their position $\psi$ on the four-sphere at fixed $p/\tilde M=0.01$. The color conventions are explained in the main text. There is a single fat-thin merger in this regime represented by the green dot. On the right, we plot the temperature of all the static M5-brane configurations as a function of their position $\psi$ at fixed $p/\tilde M = 0.03$. This is still a phase diagram in regime I.}
\label{SolTemp_1_3}
\end{figure}

\begin{figure}[t!]
\begin{center}
\includegraphics[width=7.6cm]{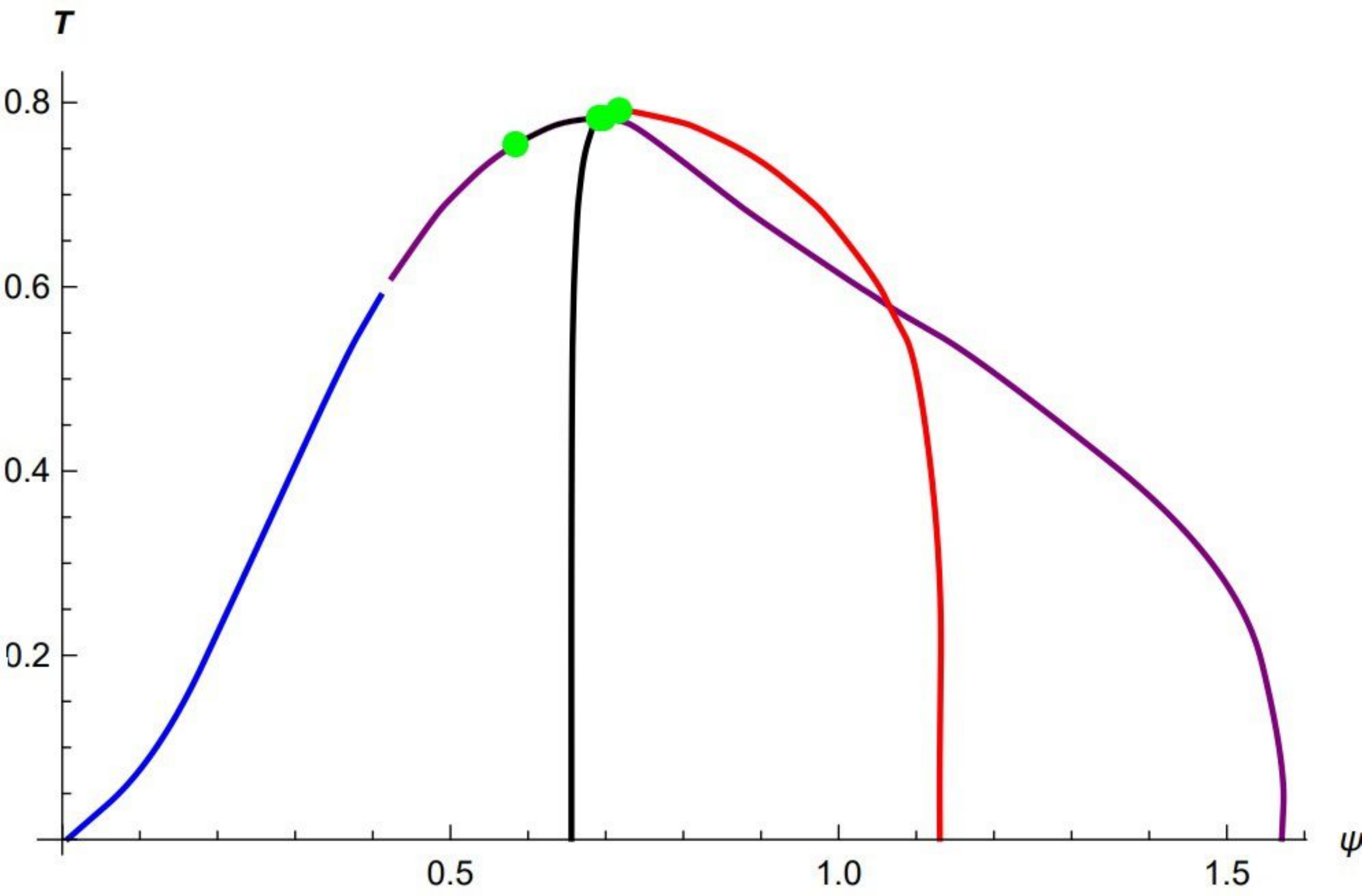}
\hspace{0.1cm}
\includegraphics[width=7.6cm]{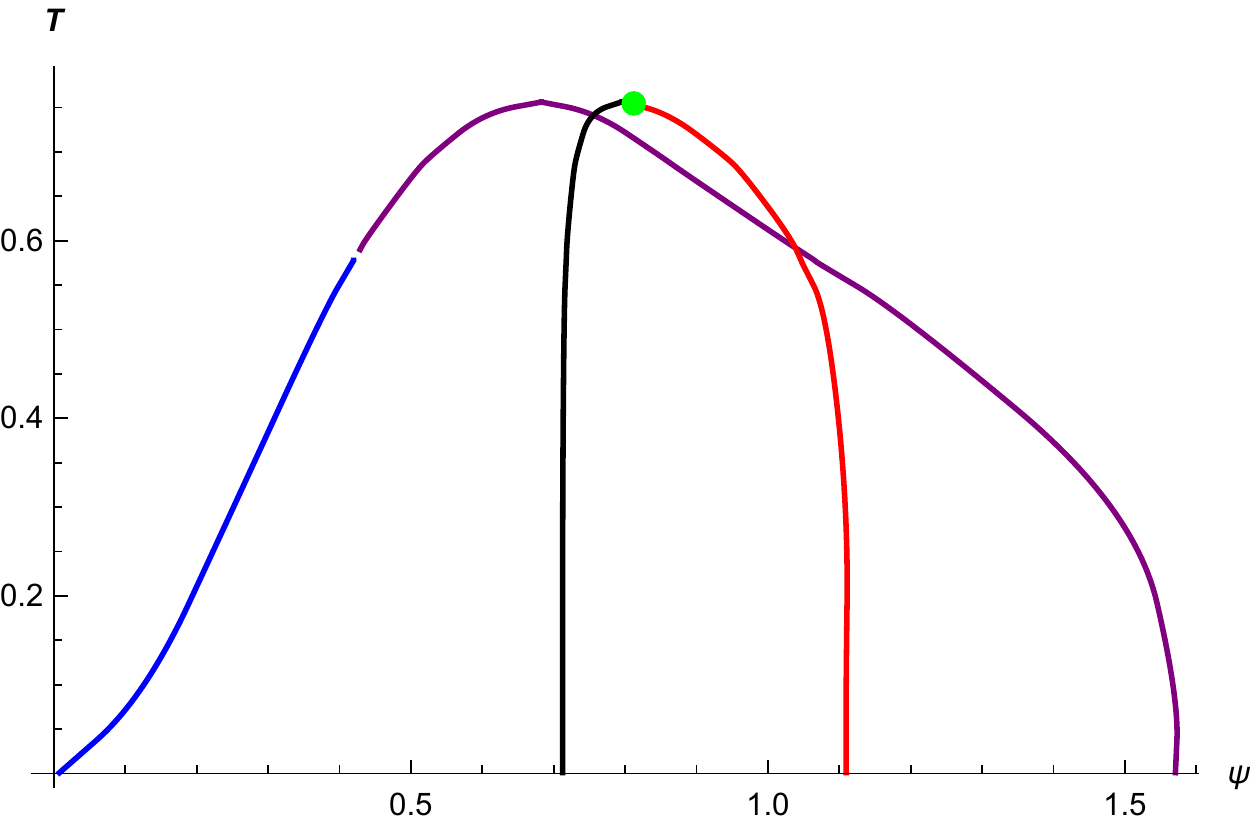}
\end{center}
\caption{On the left, we plot the temperature of all the static M5-brane configurations as a function of their position $\psi$ at fixed $p/\tilde M = 0.035$. This is a representative phase diagram in regime II. There are three merger points in this diagram represented by green dots and two separate branches of metastable states represented by black segments. On the right, we plot the temperature of all the static M5-brane configurations as a function of their position $\psi$ at fixed $p/\tilde M = 0.04$. This is a representative phase diagram in regime III.}
\label{SolTemp_35_4}
\end{figure}

An interesting alternative perspective to the thermal properties of the wrapped M5 branes arises from the analysis of the blackfold equations at fixed temperature $T$. Since the fixed-$T$ potential $V_T$ is not defined for all angles $\psi$, it is more informative to plot $\T$ as a function of $\psi$ for  the extrema of $V_T$ at each value of $p/\tilde M$. In Figs.\ \ref{SolTemp_1_3} and \ref{SolTemp_35_4} we present these plots for $p/\tilde M=0.01,0.03,0.035,0.04$. The color conventions for these plots are as follows:
\begin{itemize}
\item The blue curves represent unstable configurations of fat M5 black branes with values of $\alpha$ in the $+$ branch in \eqref{potTac}.
\item The purple curves in Figs.\ \ref{SolTemp_1_3} and Figs.\ \ref{SolTemp_35_4} represent unstable configurations in the $-$ branch in \eqref{potTac}.
\item The black curves represent metastable states. They belong to both the $+$ branch and the $-$ branch.
\item The orange curves in Figs.\ \ref{SolTemp_1_3} and the red curves in Figs.\ \ref{SolTemp_35_4} represent unstable thin M5-brane configurations. They belong to the $+$ branch.
\item The green dots represent mergers of a metastable with an unstable black hole phase. These dots are in direct correspondence with the green dots in the previous plots. Other points where different curves intersect are not merger points. At these intersections there are two black hole states with the same $T$ and $\psi$ but different values of $\alpha$.
\end{itemize}

The salient features of the temperature diagrams in Figs.\ \ref{SolTemp_1_3} and \ref{SolTemp_35_4} are the following. There are again three separate regimes I, II, III, which are the fixed-$T$ versions of the corresponding regimes in the fixed-$\alpha$ and fixed-$S$ analyses.

\vspace{0.2cm}
\noindent
{\bf Regime I: small $p/\tilde M$.}
At small enough values of $p/\tilde M$ the characteristic behaviour of the phase diagram is represented by Figs.\ \ref{SolTemp_1_3}. Let us first consider the features of the left plot of Fig.\ \ref{SolTemp_1_3}. At small values of the temperature, there are four black hole phases: the blue unstable fat M5 state, the black metastable state, the orange unstable thin M5 state and the purple unstable thin M5 state. The combined branch of the orange and purple states is in one-to-one correspondence with the unstable states represented by red dots in e.g.\ Fig.\ \ref{Valpha_3}. There is no merger in this branch. In this regime, there is only a single merger which is located on the blue-black branch. The blue-black merger is the fat-metastable merger that we noted also in Fig.\ \ref{Valpha_3} in the context of the fixed-$\alpha$ and fixed-$S$ analyses. This merger involves two states in the $+$ branch. 

As we increase $p/\tilde M$ we observe two effects, which are clearly visible in the right plot of Fig.\ \ref{SolTemp_1_3}. The first one is the appearance of intermediate purple states in the blue-black pair. Unlike the case of the left plot of Fig.\ \ref{SolTemp_1_3}, in this case the merger happens in the $-$ branch, involving a metastable state (which close to the merger is in the $-$ branch) and a state in the purple branch (which is, by default, also a state in the $-$ branch). This merger occurs now at higher values of $\T$ compared to those in regime I. The second observation is that the blue-black and red-purple branches have moved closer.

\vspace{0.2cm}
\noindent
{\bf Regime II: intermediate $p/\tilde M$.}
In the second regime, where $p/\tilde M \in (\mathfrak p_1,\mathfrak p_2)$, the phase diagram has clearly rearranged. A representative case is depicted in the left plot of Fig.\ \ref{SolTemp_35_4}. In this diagram, the main metastable branch has joined to the thin unstable (red) branch. The previous blue-purple branch has joined with the remaining purple branch at large $\psi$ through a new intermediate metastable set of states. These metastable states are represented by the black segment between the two green dots at $\psi\simeq 0.6$ and 0.7 in the left plot of Fig.\ \ref{SolTemp_35_4}. There are three merger points in this diagram in direct correspondence to the mergers observed in the left plot of Fig.\ \ref{Valpha_35_4}.

\vspace{0.2cm}
\noindent
{\bf Regime III: high $p/\tilde M$.}
Above the second critical value $p/\tilde M = \mathfrak p_2$, the phase diagram exhibits the behaviour 
represented by the right plot of Fig.\ \ref{SolTemp_35_4}. In this regime there is a single thin-thin merger represented by the green dot at the joining point of the black (metastable) and red (unstable) branches. The combined blue-purple branch has no merger points and all its states are unstable. This phase diagram is directly related to the features observed in the right plot of Fig.\ \ref{Valpha_35_4} and the right plot of Fig.\ \ref{Ventro}.

\subsection{Nature of the merger points}
\label{mergergeometry}

In the previous subsections we distinguished between different branches of solutions by characterising them as thin or fat depending on the relative size of the Schwarzschild radius and the radius of the $S^3$ that the black M5 wraps. This characterisation can be made quantitatively more specific by introducing the dimensionless ratio
\beq
\label{mergeaa}
d \equiv  \frac{ p^{1/3} \hat r_0}{\hat R_{S^3}} 
= \left( \frac{p}{\tilde M} \right)^{\frac{1}{3}} \frac{1}{\sin\psi (\cos\theta \sinh\alpha \cosh\alpha)^{\frac{1}{3}}}
~,
\eeq
where 
\beq
\label{mergeab}
\hat r_0 =r_0 \left( \frac{\CC}{\QQ_5} \right)^{\frac{1}{3}} 
\eeq
is a dimensionless quantity proportional to the local Schwarzschild radius of the black hole and
\beq
\label{mergeac}
\hat R_{S^3} = \frac{(18\pi^2)^{\frac{1}{3}}}{2\pi \ell_P} R_{S^3}
\eeq
is a dimensionless quantity proportional to the radius $R_{S^3} = m^{\frac{1}{3}}b_0 \sin\psi$ of the $S^3$ that the M5 black brane wraps at an angle $\psi$. A similar measure of black hole `fatness' was introduced in \cite{Armas:2018rsy} to describe wrapped NS5 black holes in the Klebanov-Strassler background. An analogous quantity, called $\nu$, that distinguishes between thin and fat neutral black ring solutions was introduced in \cite{Emparan:2006mm}.

\begin{figure}[t!]
\begin{center}
\includegraphics[width=7.6cm]{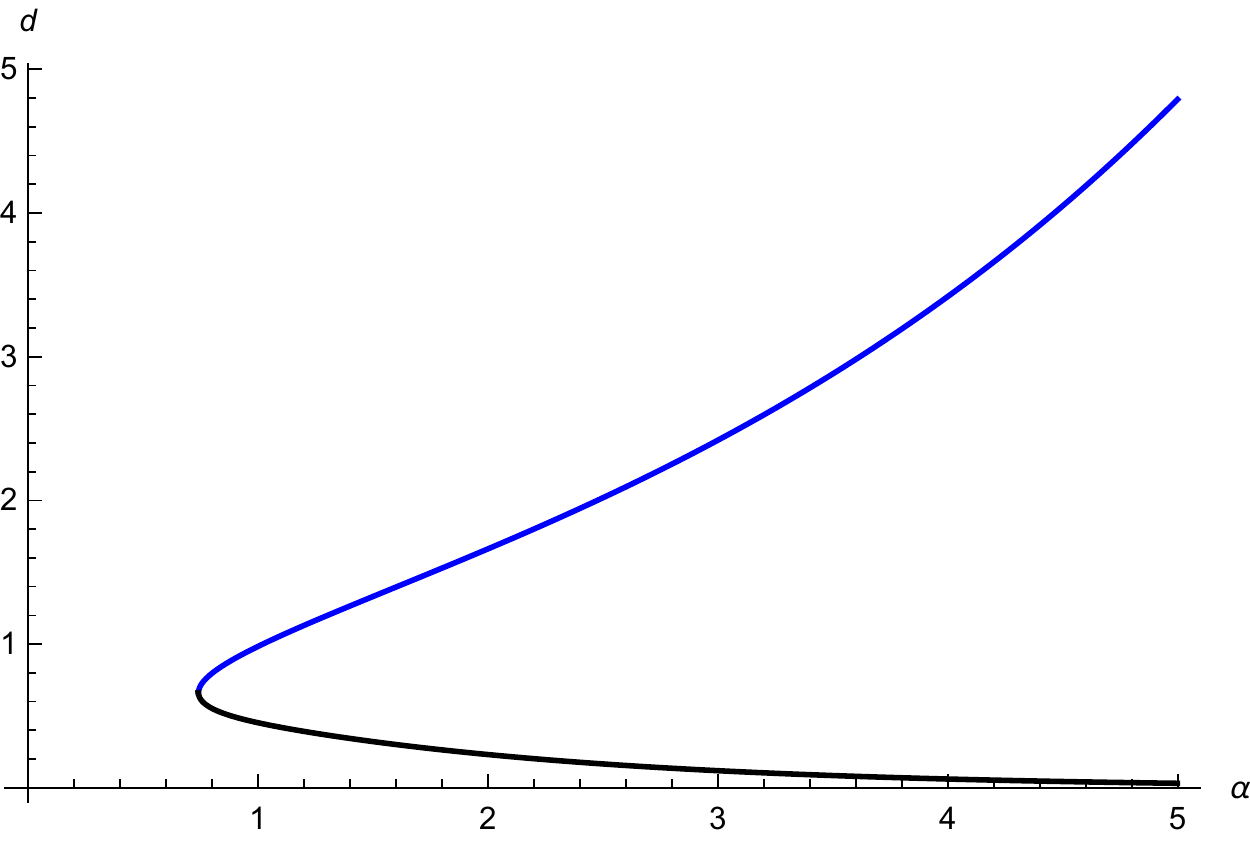}
\hspace{0.1cm}
\includegraphics[width=7.6cm]{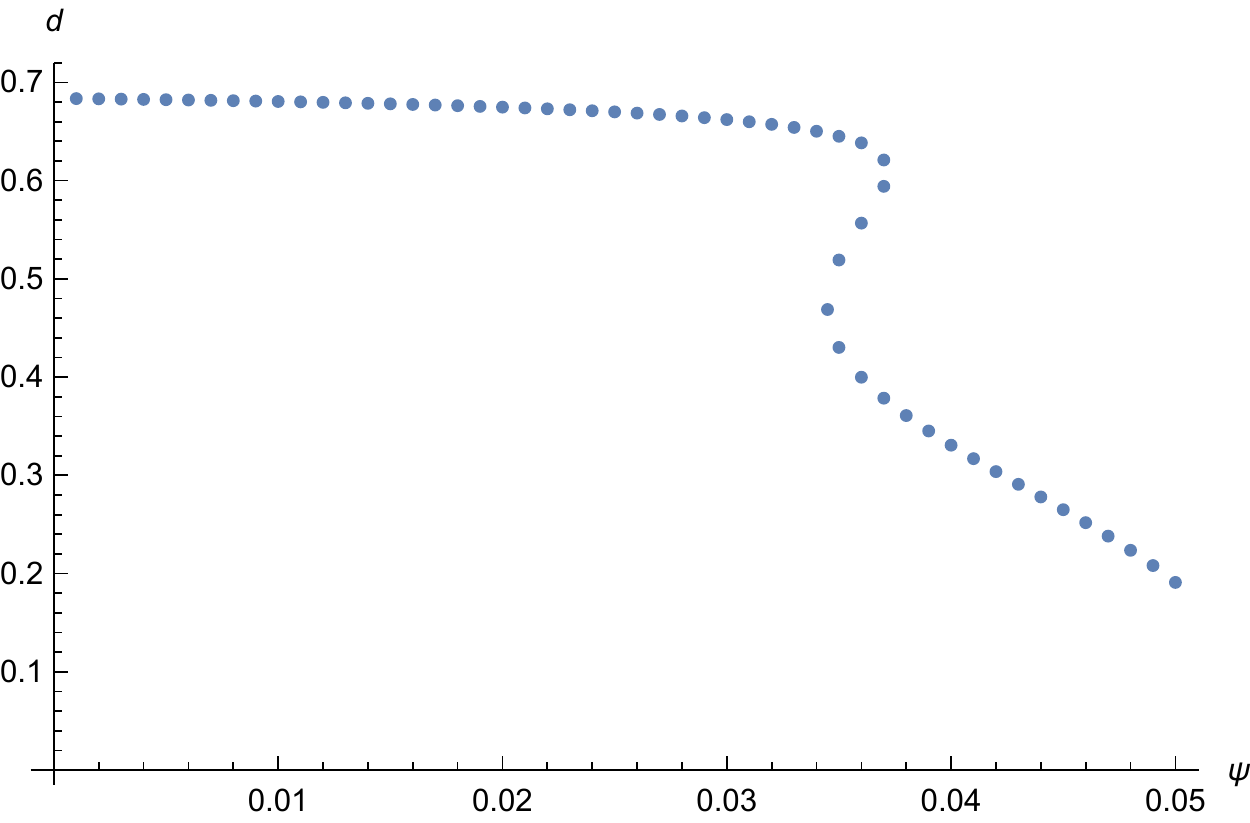}
\end{center}
\caption{On the left, we plot the fatness ratio $d$ for blue unstable and black metastable states at $p/\tilde M =0.03$. The merger of the two branches occurs at $d\simeq 1$. The near-extremal metastable configurations are thin (small values of $d$), whereas the near-extremal unstable ones are fat (large values of $d$). On the right, we plot the fatness ratio $d$ (defined in eq.\ \eqref{mergeaa}) at the merger points of the metastable states as a function of $p/\tilde M$. In region I, that involves a thin-fat merger, $d$ is almost constant. In region III, that involves a thin-thin merger, $d$ depends strongly on $p/\tilde M$. The intermediate regime II involves a multiplet of merger points.}
\label{fatthin_d1}
\end{figure}

Black hole states with $d\ll 1$ are by definition thin states where the Schwarzschild radius is comparatively smaller to the $S^3$ radius. States on the opposite part of the spectrum with $d\gg 1$ are fat states. For example, in the left plot of Fig.\ \ref{fatthin_d1} we present the value of the ratio $d$ for the blue unstable and black metastable branches at $p/\tilde M=0.03$ that are depicted by blue and black dots in Fig.\ \ref{Valpha_3}. We note that the near-extremal (i.e.\ large $\alpha$) metastable states have very low value of $d$ and are therefore thin states, whereas the corresponding unstable blue states have a very large value of $d$ and are therefore fat states. The merger occurs at a value of $d$ close to one.

According to the validity analysis of section \ref{validity}, and in particular \eqref{eq:reqm}, one finds
\beq
d\ll\left(\frac{p}{N_5}\right)^{1/3}\sinh\alpha^{-1}~~.
\eeq 
Thus, by appropriately tuning $p/N_5\gg1$ it is always possible to keep the merger points within our regimes of validity. On the other hand, as $\alpha$ increases, the validity of the unstable branch (blue curve in the left plot of Fig.~\ref{fatthin_d1}) becomes more and more restricted. 

A particularly interesting part of our discussion in this paper concerns the physics of the mergers where the metastable state is lost. The exact analysis of \cite{Cohen-Maldonado:2016cjh} shows that the existence of metastable anti-M2 states relies on the topology of the horizon. Supergravity configurations that describe point-like anti-M2s are not allowed by no-go theorems at zero temperature. At finite temperature anti-M2 black holes with spherical horizon topology are in principle allowed but require special boundary conditions for the fluxes on the horizon. Configurations of M5 black branes with non-spherical horizon topology evade these restrictions and are allowed by no-go theorems at zero temperature. As we thermalise the state, its horizon grows (namely, $d$ increases). At sufficiently high temperature one expects a transition where the horizon geometry can play a role. In that case, a scenario where the metastable state is lost would be consistent with the existing no-go theorems. 

We have already seen that the blackfold analysis verifies the expectation that the metastable state is lost at sufficiently high temperatures. The idea that the loss of the metastable state occurs because of a horizon-driven transition can be tested quantitatively by evaluating the fatness ratio $d$ at the merger points of the metastable states. Two things should happen if the above expectations are correct. Firstly, if the mergers are fat-thin mergers the transitions should occur at some value of $d$ of order 1. Secondly, some feature, e.g.\ a weak dependence of $d$ on $p/\tilde M$, should signal the dominant role of the horizon geometry. This is a highly non-trivial expectation. 

In the right plot of Fig.\ \ref{fatthin_d1} we present numerical data on $d$ evaluated at the merger points of the metastable states as a function of $p/\tilde M$. We observe that the above expectation is verified extremely well in regime I, where the metastable state is lost via a fat-thin merger. Exactly the same type of physics was observed also in the case of polarised anti-D3 branes in the Klebanov-Strassler background in \cite{Armas:2018rsy}. These results are very suggestive about the validity of the overall picture that emerges from the use of the leading order blackfold analysis.

A new feature of the M2-M5 system in CGLP compared to the D3-NS5 system in Klebanov-Strassler is the existence of regimes II and III. In Regime III the metastable state is lost in a thin-thin merger via a completely different mechanism. In this case, the metastable state does not disappear because of the horizon-related effects, but because it develops a classical instability at some critical temperature. Consistently with this picture, in the right plot of Fig.\ \ref{fatthin_d1} we observe that the values of $d$ at those mergers in regime III are much smaller and exhibit strong dependence on $p/\tilde M$. In the intermediate regime II, where multiple metastable branches occur, we observe an interesting feature of multi-valuedness in the dependence of $d$ on $p/\tilde M$. 

The co-existence of these patterns of mergers in the M2-M5 system is an interesting new prediction of the blackfold formalism. It would be very interesting to uncover further evidence for these transitions in supergravity (perhaps with numerical methods) and to understand the implications of these features in the three-dimensional QFT dual.

\section{Conclusions}
\label{conclusions}

The results of this paper, in conjunction with the exact Smarr relations in \cite{Cohen-Maldonado:2015ssa,Cohen-Maldonado:2016cjh}, provide new evidence in favor of the existence of metastable states of polarised anti-M2 branes in the CGLP background. In addition, our analysis uncovered a previously unknown pattern of black hole transitions that involve mergers of the metastable state with unstable states of wrapped M5 black branes. We presented data that support the expectation that the transition is controlled by properties of the horizon geometry when there is a merger of the metastable state with a fat M5 state. We view this observation as favourable evidence for the consistency of the proposed picture.  

Our work has led to a number of questions. We conclude with a short summary of these questions and some of the relevant open problems.

Firstly, it would be very interesting to complete the leading order blackfold analysis by constructing the full leading-order backreacted solution of wrapped M5 (black) branes in the appropriate scheme of matched asymptotic expansions. This would prove the conjecture that the constraint equations imply a regular perturbative solution of all the supergravity equations re-affirming the blackfold conjecture also in the case of anti-brane backreaction. As we noted in the main text, this exercise would not produce additional data for the leading order thermodynamic properties of the black holes of interest (beyond what has been presented in this paper), but the result could be used to determine higher derivative corrections to the blackfold equations. 

Another interesting question has to do with the stability of the solutions we described. Our analysis provides evidence (e.g.\ through the form of the fixed-entropy potential $V_S$) that under long-wavelength deformations of a certain type the metastable vacua are classically stable. It would be useful to perform a more general classical stability of the leading order blackfold equations around the metastable vacua to determine if there are other modes that can render the metastable vacua unstable. This would help clarify previous claims of instability of the metastable vacua in \cite{Bena:2014jaa,Bena:2014bxa}.

For unstable vacua, as well as semiclassically for the metastable ones, it is interesting to ask how the corresponding instabilities evolve dynamically and what is the end-state of the instability. When the solutions are extremal, there is an obvious end-point of the instability ---the supersymmetric vacuum at the south pole. For example, the metastable vacuum is expected to evolve through vacuum tunnelling to the supersymmetric state through a process that is known as brane/flux annihilation \cite{Kachru:2002gs}. In the blackfold effective description of the thermal physics we have seen that the north and south poles are strictly outside the regime of validity. However, even with this issue set aside for a moment, we notice that unlike the extremal case, the potential $V_S$ does not have any naive extrema at the two poles. Because of these features it is rather unclear what happens at the end-point of these instabilities. What kind of thermal solution, or black hole lies at the end of the evolution process of classical and semiclassical instabilities for the wrapped M5 black branes that we described?

Finally, it would be interesting to elaborate further on the dual QFT interpretation of our results. We have uncovered an intricate, previously inaccessible, pattern of thermal transitions of black hole phases in the bulk. What is the interpretation of these transitions in QFT? In this context, it would be useful to gain a more complete understanding of the nature of the finite-temperature merger points already in the gravity description. For the fat-metastable mergers, besides the features of the ratio $d$ that we reported above, we have not detected any other characteristic feature of the merger. For the thin-metastable mergers one may note the similarity with the zero-temperature case at the maximum value of the anti-brane charge where the metastable vacuum is lost. In \cite{Armas:2018rsy} we observed that this occurs close to a point where the local anti-D3 charge density vanishes. Similar observations can be made in the anti-M2 system in CGLP, but their significance is unclear.

\section*{Acknowledgements}

We would especially like to thank Thomas van Riet for useful discussions, suggestions and collaboration on a related project \cite{Armas:2018rsy}. JA is partly supported by the Netherlands Organisation for Scientific Research (NWO). The work of VN is supported by STFC under the consolidated grant ST/P000371/1. The work of NO is supported in part by the project “Towards a deeper understanding of black holes with non-relativistic holography” of the Independent Research Fund Denmark (grant number DFF-6108-00340) and Villum Foundation Experiment project 00023086.

\begin{appendix}

\section{Note on Smarr relations}  \label{smarr}
In this appendix we discuss in more detail the thermodynamic properties of stationary M2-M5 brane configurations in the CGLP background. We begin by identifying the local conserved currents that lead to global thermodynamic quantities. This can be done by requiring diffeomorphism invariance of the effective action along Killing directions as in \cite{Armas:2017pvj}, or by direct manipulation of the blackfold equations \eqref{blackeqsad} as in \cite{Armas:2012bk}. This exercise leads to the conserved currents 
\begin{equation} \label{eq:ccs}
P^a_{\boldsymbol k}=T^{a\nu}{\boldsymbol k}_\nu+\frac{1}{3!}\tilde{J}_3^{a\nu\lambda}A_{3\rho\nu\lambda}{\boldsymbol k}^\rho+\frac{1}{6!}\mathcal{J}_{6}^{a\mu_1...\mu_5}A _{6\lambda\mu_1...\mu_5}{\boldsymbol k}^\lambda~~,
\end{equation}
for some Killing vector field ${\boldsymbol k}^\mu$. For the specific static configurations that we considered in this paper, the only relevant charge obtainable from this current is the total energy
\beq
E=\int_{\mathcal{B}_5}dV_{5}P^a_{\boldsymbol{k}}u_a=\int_{\mathcal{B}_5}\left(dV_{(p)}\varepsilon-\mathcal{Q}_5\tilde{\mathbb{P}}[A_6]\right)~~,
\eeq
where $dV_{5}$ is the volume form on $\mathcal{B}_5$ and $\tilde{\mathbb{P}}[A_6]=\mathbb{P}[A_6]/dt$. We notice that this expression is the same as (minus) the Wick rotated effective action at constant entropy \eqref{thermoag}, given that $\mathbb{P}^{||}[A_3]$ vanishes for these configurations as well as the contribution of the second term in \eqref{eq:ccs}. Thus, as earlier advertised, minimising \eqref{thermoag} is minimising the total energy $E$. 

In order to proceed further, we must identify the remaining thermodynamic quantities of relevance. These are the global chemical potentials $\Phi_{\text{H}}^5$ and $\Phi_{\text{H}}^2$, which are the thermodynamic conjugates of $\mathcal{Q}_5$ and $\mathbb{Q}_2$ respectively. As such, they are derivable from the effective action \eqref{thermoag} itself according to
\beq\label{eq:chem}
\Phi_{\text{H}}^5=-\frac{\partial \mathcal{S}_S^{E}}{\partial \mathcal{Q}_5}\bigg |_{S,\mathbb{Q}_2}~~,~~\Phi_{\text{H}}^2=-\frac{\partial \mathcal{S}_S^{E}}{\partial \mathbb{Q}_2}\bigg |_{S,\mathcal{Q}_5}~~,
\eeq
where the superscript $E$ denotes the fact that we performed a Wick rotation in \eqref{thermoag} and integrated over the time circle with size $1/T$. Using \eqref{eq:chem} we find
\beq
\begin{split}
\Phi_{\text{H}}^5=\int_{\mathcal{B}_5}dV_5 |\boldsymbol{k}| \left(\Phi_5-\frac{\Phi_2}{\text{Vol}_\perp}\int_{\mathcal{M}_3^\perp}\mathbb{P}^\perp[A_3]\right)-\int_{\mathcal{B}_5}\tilde{\mathbb{P}}[A_6]~~,~~\Phi_{\text{H}}^2=\int_{\mathcal{B}_5}dV_5 |\boldsymbol{k}| \frac{\Phi_2}{\text{Vol}_\perp}~~,
\end{split}
\eeq
where $\text{Vol}_\perp$ is the volume of $\mathcal{M}_3^\perp$ and we again used that $\mathbb{P}^{||}[A_3]=0$. The form of $\Phi_{\text{H}}^2$, in particular, is the expected form of the global chemical potential of a higher-form fluid \cite{Armas:2018ibg}. Using all thermodynamic quantities, one may construct the Gibbs free energy
\beq \label{eq:gibbsf}
I_{\text{E}}=E-TS-\Phi_{\text{H}}^5\mathcal{Q}_5-\Phi_{\text{H}}^2\mathbb{Q}_2=\frac{\Omega_4}{16\pi G}\int_{\mathcal{B}_5}dV_5 |\boldsymbol{k}| r_0^3=\frac{TS}{3}~~,
\eeq
and from here we derive the Smarr relation for polarised M2 branes into M5 branes
\beq
E=\frac{4}{3}TS+\Phi_{\text{H}}^5\mathcal{Q}_5+\Phi_{\text{H}}^2\mathbb{Q}_2~~,
\eeq
which agrees with the corresponding Smarr relation derived in \cite{Cohen-Maldonado:2016cjh}. The definition of Gibbs free energy \eqref{eq:gibbsf} allows for the construction of an effective action at constant $T,\Phi_{\text{H}}^5, \Phi_{\text{H}}^2$, which takes the form
\beq
\mathcal{S}_{\mathcal{G}}=-\int_{\mathcal{M}_6}\sqrt{-\gamma}\mathcal{G}~~,~~\mathcal{G}=\frac{\Omega_4}{16\pi G}r_0^3~~.
\eeq
Minimising $\mathcal{S}_{\mathcal{G}}$ implies the first law of black hole thermodynamics $dE=TdS+\Phi_{\text{H}}^5d\mathcal{Q}_5+\Phi_{\text{H}}^2d\mathbb{Q}_2$.

\end{appendix}

\renewcommand{\tt}{\normalfont\ttfamily}
\providecommand{\href}[2]{#2}\begingroup\raggedright\endgroup


\end{document}